\documentclass[fleqn,usenatbib]{mnras}

\usepackage{newtxtext,newtxmath}

\usepackage[T1]{fontenc}

\DeclareRobustCommand{\VAN}[3]{#2}
\let\VANthebibliography\thebibliography
\def\thebibliography{\DeclareRobustCommand{\VAN}[3]{##3}\VANthebibliography}

\usepackage{graphicx}	
\usepackage{amsmath}	
\usepackage{amsfonts}
\usepackage{float}
\usepackage{subcaption} 
\usepackage{siunitx}    
\usepackage{wrapfig}
\usepackage{adjustbox}

\title[M87* constraints on hairy Kerr BHs]{Parameter estimation of hairy Kerr black holes from its shadow and constraints from M87*}

\author[M. Afrin, R. Kumar and S.G. Ghosh]{
Misba Afrin,$^{1}$\thanks{E-mail: me.misba@gmail.com}
Rahul Kumar,$^{1}$
and Sushant~G.~Ghosh$^{1,2}$
\\
$^{1}$Centre for Theoretical Physics, Jamia Millia Islamia, New Delhi 110025, India\\
$^{2}$Astrophysics and Cosmology Research Unit, School of Mathematics, Statistics and Computer Science, University of KwaZulu-Natal, Private Bag 54001,\\ Durban 4000, South Africa
}

\date{Accepted 2021 April 29. Received 2021 April 28; in original form 2021 April 2}

\pubyear{2021}

\begin{document}
\label{firstpage}
\pagerange{\pageref{firstpage}--\pageref{lastpage}}
\maketitle

\begin{abstract}
The recently obtained hairy Kerr black holes, due to additional sources or surrounding fluid, like dark matter, with conserved energy-momentum tensor, have a deviation $\alpha$ and primary hair $l_0$, apart from rotation parameter $a$ and mass $M$. In the wake of the \textit{Event Horizon Telescope} (\textit{EHT}) observations of the supermassive black hole M87*,  a recent surge in interest in black hole shadows suggests comparing the black holes in general relativity (GR) and modified theories of gravity (MoG) to assess these models' differences.  Motivated by this, we take on an extensive study of the rotating hairy Kerr black holes, which encompasses, in particular cases, the Kerr black hole ($\alpha=0$). We investigate ergosphere and shadows of the black holes to infer that their size and shape are affected due to the $l_0$ and are found to harbour a richer chaotic structure. In particular, the hairy Kerr black holes possess smaller size but more distorted shadows when compared with Kerr black holes.  We also estimate the parameters $l_0$ and $a$ associated with hairy Kerr black holes using the shadow observables. The inferred circularity deviation $\Delta C \leq 0.1$ for the M87* black hole is satisfied, whereas shadow angular diameter $\theta_{d}=42 \pm 3 \mu as$, within $1 \sigma$ region, for a given choice of $\alpha$, places bounds on the parameters $a$ and $l_0$. Interestingly, the shadow axial ratio obeying $1< D_x \lesssim 4/3$ is in agreement with the \textit{EHT} results and thus eventuates in the hairy Kerr black holes being suitable candidates for astrophysical black holes.
\end{abstract}




\section{Introduction}\label{Intro}
Recently, the international \textit{Event Horizon Telescope} (\textit{EHT}) collaboration has unveiled the first shadow image of a supermassive black hole M87* with angular gravitational radius $3.8 \pm 0.4$ $\mu$as -- an asymmetric bright emission ring with a diameter of $42 \pm 3$ $\mu$as, exhibiting a deviation from circularity $\Delta C \leq 0.1$ and an axis ratio $\lesssim 4/3$, and it is consistent with the shadow of a Kerr black hole in general relativity (GR). The black hole shadows have become a physical reality with the detection of the M87* black hole's horizon-scale image \citep{Akiyama:2019cqa,Akiyama:2019bqs,Akiyama:2019eap}. The bright, sharp photon ring, a projection along the photons' null geodesics orbiting around the black hole, encompasses the shadow and explicitly depends on the black hole parameters \citep{Johannsen:2010ru}. It has been found that the quantitative features are not sufficient to distinguish between black holes using different theories of gravity. Thus, using the M87* black hole shadow, one can investigate the viability of black holes in modified theories of gravity (MoG) in explaining the observational data while placing constraints on the black hole parameters and the \textit{EHT} observations thus became an important tool to test strong gravitational fields. 

The shadows of rotating black holes attracted a lot of attention in the past few decades, led by the seminal work by \cite{Synge:1966okc} and  \cite{Luminet:1979nyg}, who formulated the expression for the angular radius of the photon capture region around the Schwarzschild black hole. Later, \cite{Bardeen:1973tla} analysed the shadow of the Kerr black hole and showed that the spin would cause distortion in the shape of the shadow. Applications of shadow in unravelling the useful information regarding the near horizon features of gravity have aroused a flurry of activities in the analytical investigations, and numerical simulation of shadows for black holes in GR \citep{Falcke:1999pj,Vries2000TheAS,Shen:2005cw,Yumoto:2012kz,Atamurotov:2013sca,Papnoi:2014aaa,Abdujabbarov:2015xqa,Atamurotov:2015xfa,Cunha:2018acu,Kumar:2018ple}. The shadows of black holes in  MoG have been found to be smaller and more distorted when compared with the Kerr black hole shadow \citep{Amarilla:2010zq,Amarilla:2011fx,Amarilla:2013sj,Amir:2017slq,Singh:2017vfr,Mizuno:2018lxz,Ghosh:2020ece,Kumar:2020owy}. An extensive work on shadows in higher dimensions can be found in literature \citep{Amir:2017slq,Eiroa:2017uuq,Vagnozzi:2019apd,Banerjee:2019nnj}. 
 Also, the shadow of black holes in GR coupled to Non-linear electrodynamics has been used to extract upper limits on the black hole magnetic charge \citep{Allahyari:2019jqz}. One of the motivations behind the multifarious works on black hole shadow is that the trajectories of light near black holes are related to the background geometry's essential features and properties. Moreover, by observing the size and deformation of shadow, the spin, mass parameter and possibly other global charges or \textit{hair}-parameters of the black holes can be calculated \citep{Hioki:2009na,Tsupko:2017rdo,Cunha:2019dwb,Cunha:2019ikd,Kumar:2018ple,Khodadi:2020jij}. Besides, it is also useful in testing theories of gravity \citep{Kramer:2004hd}. 

Observations of the shadow of supermassive black holes M87* allow us to study the properties of black holes and the nature of strong-field gravity \citep{Akiyama:2019cqa,Akiyama:2019bqs,Akiyama:2019eap}. In principle, one can use black hole shadow observations to place constraints on the hairy Kerr black hole, which can also be a solution to MoG that do not admit the Kerr solution. We here investigate, whether black hole shadow observations can constrain hairy Kerr black holes \citep{Contreras:2021yxe}, which has additional deviation parameter $\alpha$ and primary hair $l_0$ from the Kerr black hole.  Furthermore, we investigate the horizons, ergoregions and shadow cast by the hairy Kerr black holes \citep{Contreras:2021yxe}. Furthermore, we obtain the null geodesic equations of motions in the first-order differential form using the Hamilton-Jacobi approach, and the analytical expressions for the photon region are thus determined. An investigation of the deviation parameters' effects shows that the hairy Kerr black holes cast smaller and more distorted shadows than the Kerr black hole. We use shadow observables to characterize the shadow's shape, size and distortion, and estimate the black hole parameters $a$ and $l_0$ thereof. The scenario in this paper is that we consider M87* as hairy Kerr black holes, and use the results of the \textit{EHT} observations to constrain the parameters of hairy Kerr black holes.

The hairy Kerr black hole that we are interested in, is obtained using the gravitational decoupling (GD) approach \citep{Contreras:2021yxe}. It has a source satisfying the strong energy condition (SEC) and gives an extended Kerr metric which was termed as Kerr black holes with primary hair by Contreras \textit{et. al.} \citep{Contreras:2021yxe}.
Hairy black holes are termed as stationary black hole solution with new global charges, which are not associated with Gauss law \citep{Herdeiro:2015waa}, e.g., black holes with scalar hair \citep{Herdeiro:2014goa,Gao:2021luq}, or proca hair \citep{Herdeiro:2016tmi}. See Ref. \citep{Herdeiro:2015waa} for a recent review on black holes with hair due to global charge.

The  hairy Kerr black holes, in question, may arise due to surrounding fluid-like dark matter, one of the essential open fundamental questions of physics. Indeed, dark matter constitutes 25 per cent of the universe's energy density \citep{Tanabashi:2018oca}.  Numerous astronomical questions are likely to be addressed with the \textit{EHT} observations while we can also expect new exciting questions. However, it is also interesting to inquire whether the \textit{EHT} data on M87* could also shed light on the properties of dark matter.

This paper is organized as follows. In Section~\ref{Sec2}, we consider hairy black holes, discuss the effect of $l_0$, for given choice of $\alpha$, on horizon structure and ergoregions. Section~\ref{Sec3} is devoted to a brief review of the null geodesic equations in the hairy Kerr black hole spacetimes. The impact of the hair parameter $l_0$  on the black hole shadow is the subject of Section~\ref{Sec4}. We present the observables for shadow characterization and use them to estimate the parameters associated  with hairy Kerr black holes in Section~\ref{Sec5}. Assuming M87* as hairy Kerr black holes, we calculate the circularity deviation, angular diameter and axis ratio observables over the entire parameter space for $90\si{\degree}$ and $17\si{\degree}$ inclination angles in Section~\ref{Sec6}. Finally, we summarize our main findings in Section~\ref{Sec7}.  We use geometrized units $G=1$, $c=1$, unless units are specifically defined.

\section{Hairy Kerr back holes }\label{Sec2}
The GD approach is precisely designed to find deformation of the known solution of GR due to the additional surrounding sources like dark matter or dark energy \citep{Contreras:2021yxe} (see also \citep{Ovalle:2017fgl,Ovalle:2019qyi}). The GD approach leads to deformed or hairy Schwarzchild black holes \citep{Contreras:2021yxe} given by 
\begin{align}\label{metric1}
ds^2=-&\left[1-\frac{2M}{r}+\alpha e^{{-r}/(M-\frac{l_{0}}{2})}\right]dt^2+r^2(d\theta^2+\sin^2{\theta}d\phi^2)\nonumber\\
+&\frac{1}{\left[1-\frac{2M}{r}+\alpha e^{{-r}/(M-\frac{l_{0}}{2})}\right]}dr^2\ .
\end{align}
Here, $M$ is the black hole mass, $\alpha$ is the deviation parameter and $l_0\leq2M$, corresponds to the primary hair which determines asymptotic flatness. The surrounding matter that leads to deformation of the Schwarzchild black holes (\ref{metric1}) is described by the conserved stress-energy tensor $S_{\mu\nu}$ that satisfies the SEC. The hairy black hole (\ref{metric1}) encompasses the Schwarzchild black hole in the absence of the surrounding matter ($\alpha=0$). The rotating black hole solutions in MoG are essential as they offer an arena to test these theories through astrophysical observation. There is hardly any test for non-rotating solutions since the black hole spin is crucial in any astrophysical process.
\begin{figure}
\includegraphics[scale=0.55]{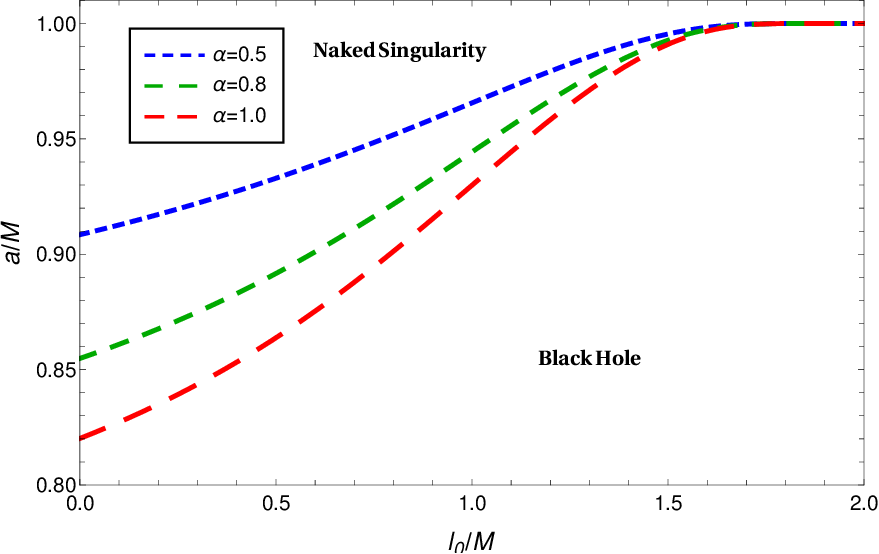}
\caption{\label{fig1} The parameter space ($l_0-a$) of the hairy Kerr black hole. The lines corresponding to extremal black holes separate black holes from naked singularities. }
\end{figure}
\begin{figure*}
\begin{tabular}{c c}
    \includegraphics[scale=0.55]{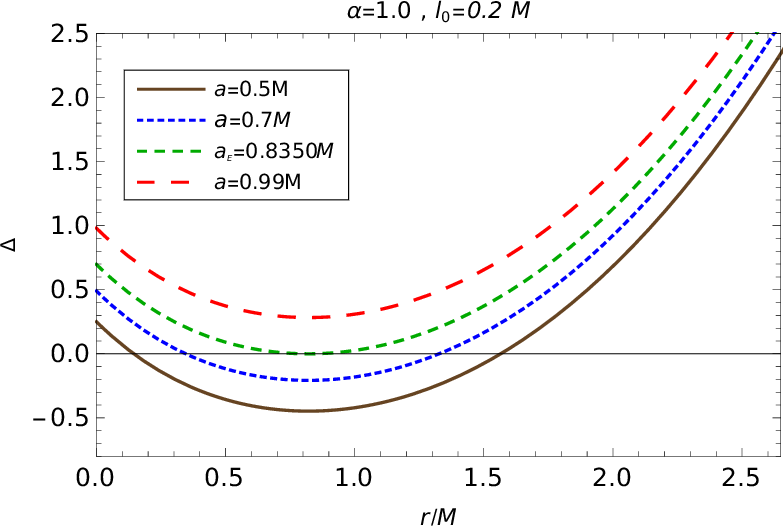}&
    \includegraphics[scale=0.55]{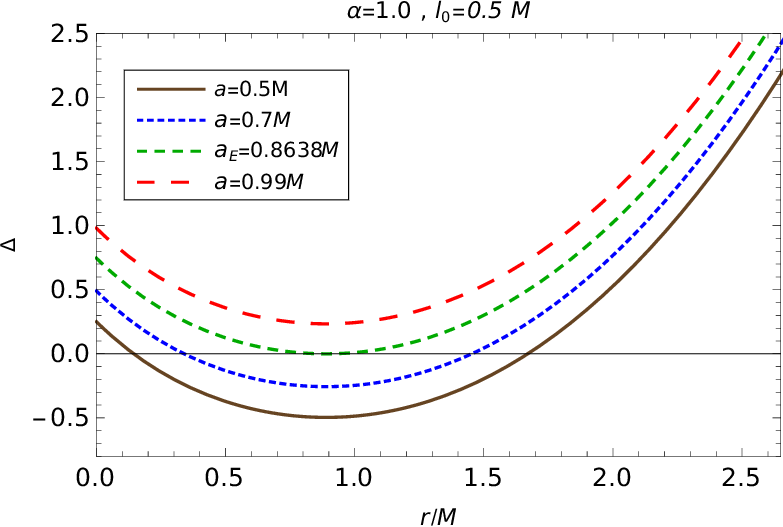}\\
     \includegraphics[scale=0.55]{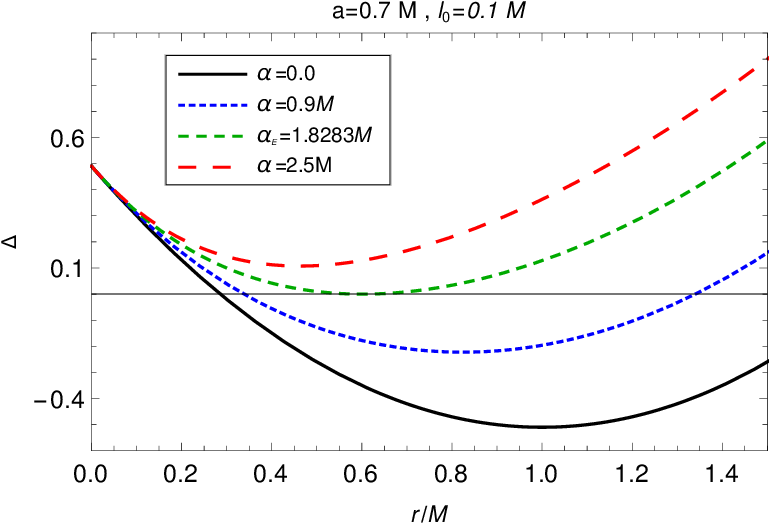}&
    \includegraphics[scale=0.55]{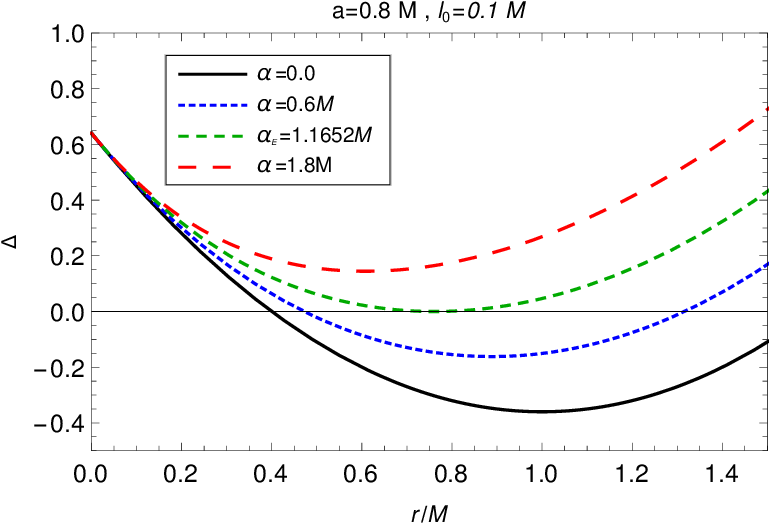}\\
    \includegraphics[scale=0.55]{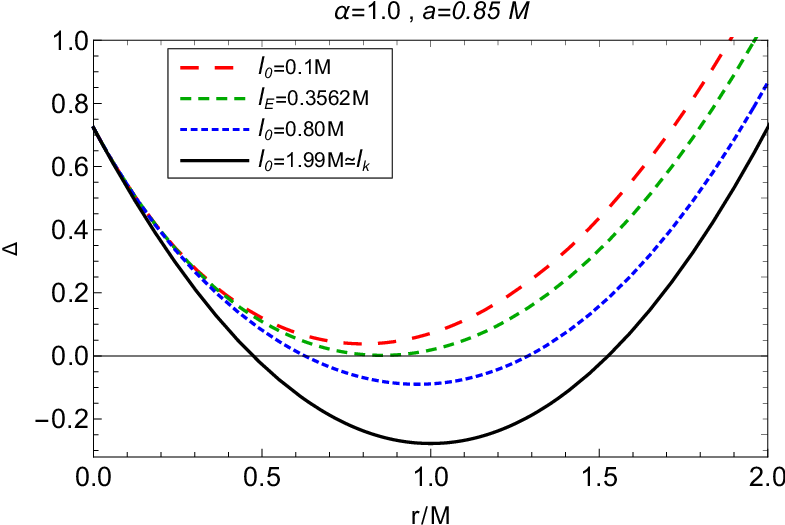}&
    \includegraphics[scale=0.55]{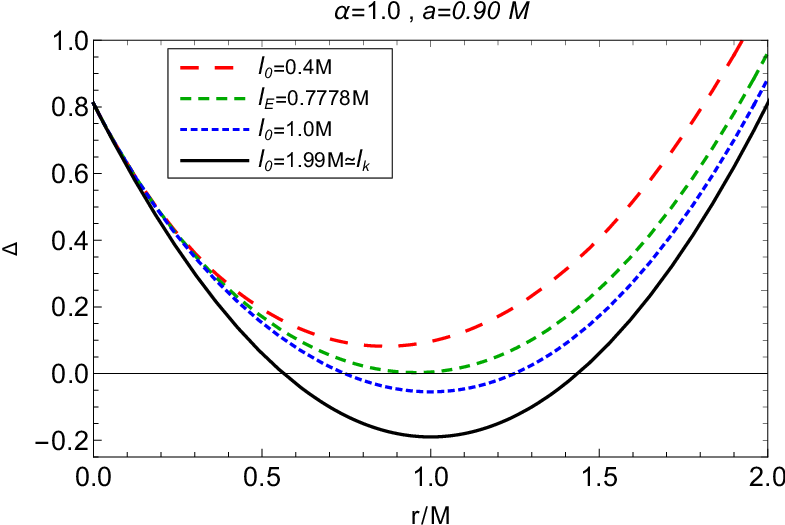}
\end{tabular}
\caption{\label{fig2}Horizons of the hairy Kerr black holes are shown with varying $a$ [top], $\alpha$ [middle] and $l_0$ [bottom] and compared with the Kerr black holes ($\alpha=0$ or $l \to l_k$).}
\end{figure*}
\begin{figure*}
	\centering
\begin{tabular}{c c c c}
	\includegraphics[scale=0.44]{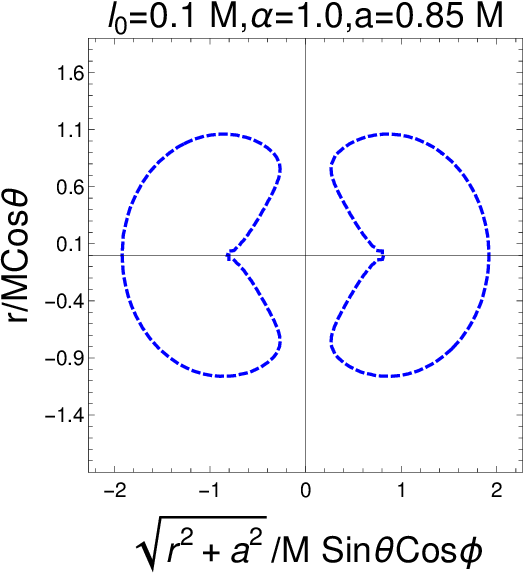}&
	\includegraphics[scale=0.44]{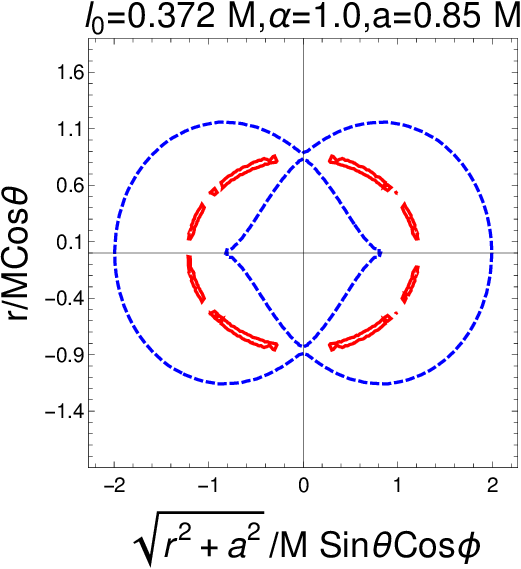}&
	\includegraphics[scale=0.44]{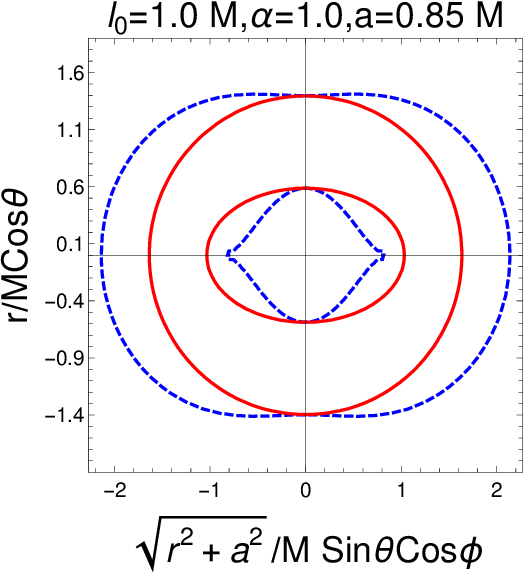}&
	\includegraphics[scale=0.44]{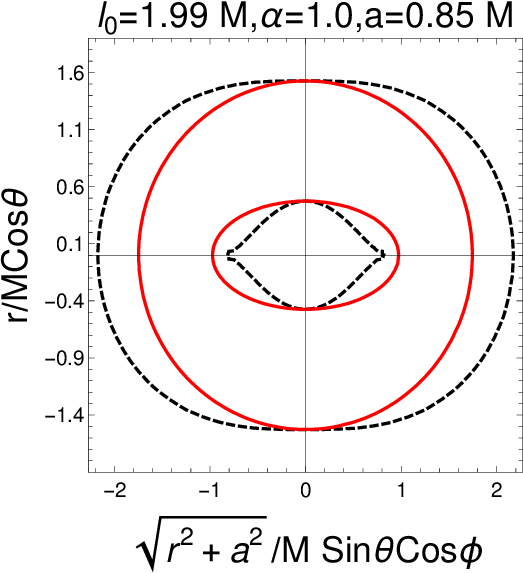}\\
	\includegraphics[scale=0.44]{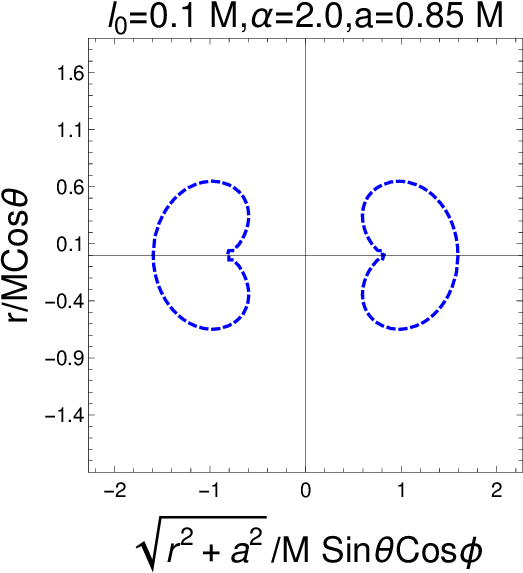}&
	\includegraphics[scale=0.44]{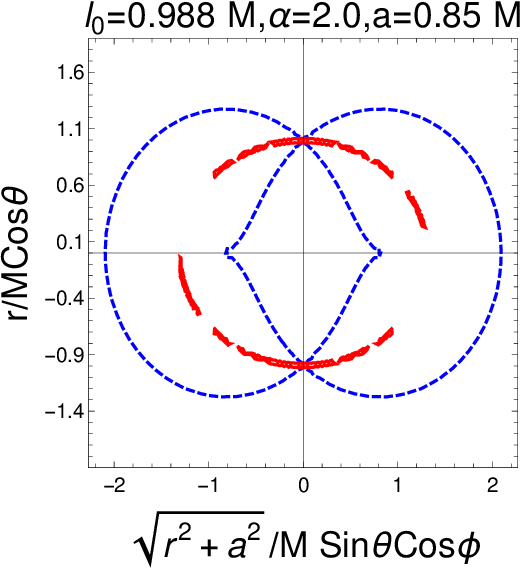}&
	\includegraphics[scale=0.44]{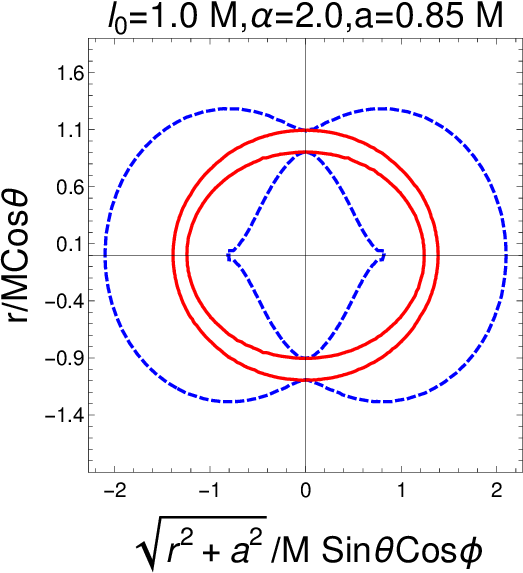}&
	\includegraphics[scale=0.44]{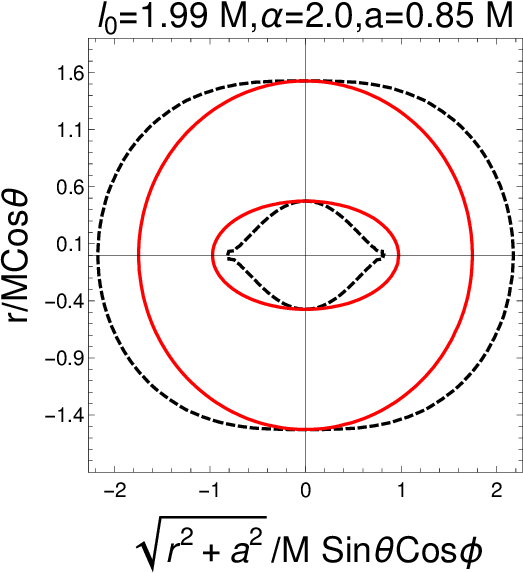}
\end{tabular}
\caption{The cross-section of event horizon (outer red line), SLS (outer blue dotted line) and ergoregion  of hairy Kerr black holes. The black dotted curves correspond to event horizon of Kerr black hole ($l_0 \to  l_k$)}\label{ergo}	
\end{figure*}

Hence, we consider the rotating counterpart of the  spherically symmetric solution.
The stationary and axisymmetric counterpart of the spherically symmetric solution (\ref{metric1}) described by parameters $M$, $a$, $l_0$ and $\alpha$, in Boyer-Lindquist coordinates reads \citep{Contreras:2021yxe}
\begin{align}\label{metric}
ds^{2}=-&\left[\frac{\Delta-a^{2}\sin^{2}\theta}{\Sigma}\right]dt^{2}-2a\sin^2{\theta}\left[1-\frac{\Delta-a^2\sin^2{\theta}}{\Sigma}\right]dtd\phi\nonumber\\
+& \sin^{2}\theta \left[\Sigma+a^2\sin^2{\theta}\left(2-\frac{\Delta-a^2\sin^2{\theta}}{\Sigma}\right)\right]d\phi^{2}+\frac{\Sigma}{\Delta}dr^{2}\nonumber\\
+&\Sigma d\theta^{2},
\end{align}
with $\Delta=r^{2}+a^{2}-2Mr+\alpha r^{2}e^{{-r}/(M-\frac{l_{0}}{2})}$ and  $\Sigma=r^{2}+a^{2}\cos^{2}\theta$, where $a$ denotes the angular momentum. The $\alpha$ is a generic parameter that measures potential deviation of metric (\ref{metric}) from the standard Kerr black holes and is related to $l_0$ via $l_0=\alpha l$. The black hole (\ref{metric}) is termed as hairy Kerr black holes \citep{Contreras:2021yxe} which encompasses the Kerr black hole in the absence of surrounding matter ($\alpha=0$). The solution (\ref{metric}) results from the surrounding matter described by stress tensor $S_{\mu \nu}$ satisfying the SEC. Thus metric (\ref{metric}) can be seen as a prototype non-Kerr black hole and with an additional deviation parameter $\alpha$ and primary hair $l_0$ \citep{Contreras:2021yxe}. When written in Boyer-Lindquist coordinate, it is same as Kerr black hole with $M$ replaced by $m(r)=M-\frac{\alpha}{2}r e^{-r/(M-\frac{l_0}{2})}$. In what follows, we shall investigate how primary hair $l_0$ affects the horizons and shadows. 

The metric (\ref{metric}) is singular at spacetime points where $\Sigma\neq0$ and $g^{\alpha\beta}\partial_{\alpha}r\partial_{\beta}r=g^{rr}=\Delta=0$, which corresponds to the event horizon. Thus the zeros of
\begin{eqnarray}\label{horizon}
r^{2}+a^{2}-2Mr+\alpha r^{2}e^{{-r}/(M-\frac{l_{0}}{2})}=0,
\end{eqnarray}
give the black hole horizons: $r_+$ denotes outer (event) horizon and $r_-$ the inner (Cauchy) horizon. Equation (\ref{horizon}), when $\alpha=0$ (Kerr black hole), admits exact solutions
\begin{eqnarray}
r_{\pm}^{k} = M\pm\sqrt{M^2-a^2}.
\end{eqnarray}
Here $r_{\pm}^{k}$ corresponds to the horizons associated with the Kerr black hole. To ensure the event horizon's existence, we require $M \geq a$. It defines a parameter space region, where metric represents a black hole and not a naked singularity. The case $M < a$, contains a naked singularity. It is clear that the metric (\ref{metric}) describes non-extremal black hole for $r_+>r_-$ and when $r_+=r_-$, one obtains an extremal black hole.  

Numerical analysis of equation~(\ref{horizon}) reveals that depending on the values of $M$, $a$, $\alpha$ and $l_{0}$ there can exist a maximum of two distinct real positive roots, or equal roots, or no-real positive roots. They respectively correspond to the non-extremal black holes with Cauchy and event horizons, extremal black holes, and no-black holes scenarios. The degenerate condition of the horizons gives a bound on the black hole parameters $l_0$ and $a$ requiring them to be restricted to a region where the metric (\ref{metric}) represents a black hole and not a naked singularity and the ($l_0$-$a$) parameter space thus obtained is shown in Fig.~\ref{fig1}. 

From the horizon structure of the hairy Kerr black holes depicted in Fig.~\ref{fig2}, it turns out that, for given values of $\alpha$ and $l_0$, there exists critical extremal value of $a$, $a_E$ and likewise there exists critical extremal value of $l_0$, $l_E$ for given value of $\alpha$ and $a$, where $\Delta=0$ has a double root. The hairy Kerr black hole exists when $a<a_E$ while for $a>a_E$ there is a naked singularity and the $a_E$ depends on $l_0$, e.g., for $l_0=0.2M, 0.5M$ respectively, $a_E=0.8350M, 0.8638M$. Also $l_0>l_E$ corresponds to hairy Kerr black holes with Cauchy and event horizons whereas $l_0<l_E$ implies existence of a naked singularity. It may be noted that, for $a=0.85M, 0.90M$, $l_E=0.3562M, 0.7778M$ respectively. Indeed $a_E$ increases with the increase in $l_0$ likewise $l_E$ increases with increase in $a$. For fixed values of $M$, $a$ and $l_0$, the hairy Kerr black holes have decreasing event horizon radius and increasing Cauchy horizon radius  with the increase in deviation parameter $\alpha$ from the Kerr limit, $\alpha=0$ (cf. Fig.~\ref{fig2}). Thus the hairy Kerr black holes have smaller event horizon radii than the Kerr black hole.  At the extremal value  $\alpha=\alpha_E$, the two horizons coincide and the radius of this extremal horizon ($r_E$) increases with the increase in rotational parameter $a$ while the value of $\alpha_E$ decreases, for e.g.,  $\alpha_E=1.8283$, $r_{E}=0.6009 M$ for $a=0.7 M$ and $\alpha_E=1.1652$, $r_{E}=0.7615 M$ for $a=0.8 M$ (cf. Fig.~\ref{fig2}).

On the other hand, at the static limit surface (SLS), the asymptotic time-translational Killing vector $\chi^{i}=(\frac{\partial}{\partial t})^{i}$, becomes null, i.e.,
\begin{eqnarray}\label{SLS}
   \chi^{i}\chi_{i}=g_{tt}=r^{2}+a^{2}\cos^{2}{\theta}-2Mr+\alpha r^{2}e^{{-r}/(M-\frac{l_{0}}{2})}=0.
\end{eqnarray}
Thus, the radial coordinates of SLS are the real positive roots ($r_{SLS}^{\pm}$) of equation~(\ref{SLS}) which admits two possible solutions. The larger of the two roots corresponds to the outer SLS, denoted by $r_{SLS}^{+}$.  The analysis of the zeros of equation~(\ref{SLS}), for a given values of $a$, $l_{0}$ and $\theta$, disseminate a critical parameter $\alpha_{SLS}^{E}$ such that equation~(\ref{SLS})  has no root if $\alpha > \alpha_{SLS}^{E}$, a double root at $\alpha = \alpha_{SLS}^{E}$, and two simple roots if $\alpha < \alpha_{SLS}^{E}$ .
The ergoregion bounded between $r_{+} < r < r_{SLS}^{+}$, where the timelike killing vector $\chi^{i}$ becomes spacelike, and an observer necessarily follows the worldline of $\chi^{i}$, has been shown in Fig.~\ref{ergo} for the hairy Kerr black hole (\ref{metric}). The parameter $l_0$  has a diminishing effect on the ergoregion's size, as the ergoregion gets smaller with the increase in $l_0$ for fixed values of the parameters $\alpha$ and $a$. Thus, the hairy Kerr black holes have larger ergoregions than those for the Kerr black hole ($l=l_k$). Moreover, in the non-Kerr limit ($l_0 \neq l_k$), the increase in the $\alpha$  increases the size of the ergoregion at fixed values of $l_0$ and $a$. It is also worthwhile noting that the decrease in the parameter $l_0$ leads to disconnected event horizons for fast rotating hairy Kerr black holes (cf. Fig.~\ref{ergo}). The ergoregion acquired its name owing to the theoretical possibility of extracting energy and mass from this region via the Penrose process \citep{Penrose:1971uk}.

The frame dragging effect of the hairy Kerr black hole becomes evident from the off diagonal elements of metric (\ref{metric}) viz., $g_{t\phi}$. Due to this effect, a stationary observer outside the event horizon, moving with zero angular momentum with respect to an observer at spatial infinity can rotate with the black hole with an angular velocity given by \citep{Pugliese:2018hju}
\begin{eqnarray}\label{omega1}
    \tilde{\omega}=\frac{d\phi}{d t}=-\frac{g_{t\phi}}{g_{\phi\phi}}=\frac{2ar(M-\alpha \frac{r}{2}e^{{-r}/(M-\frac{l_{0}}{2})})}{(r^2+a^2)^2-a^2 \Delta\sin^2{\theta}},
\end{eqnarray}
which monotonically increases as the observer approaches the black hole and ultimately at the event horizon, where the observer begins co-rotating with the black hole, takes the maximum value:
\begin{eqnarray}\label{omega2}
    \Omega= \tilde{\omega} |_{r=r_{+}}=\frac{2ar_{+}(M-\alpha \frac{r_{+}}{2}e^{{-r_{+}}/(M-\frac{l_{0}}{2})})}{(r_{+}^2+a^2)^2}.
\end{eqnarray}
Here $\omega$ is the black hole angular velocity, which in the limit $\alpha=0$ reads
\begin{eqnarray}
    \Omega_{Kerr}=\frac{2Mar_{+}}{(r_{+}^2+a^2)^2}
\end{eqnarray}
and corresponds to the angular velocity of Kerr black hole \citep{Frolov:2014dta}. It is clear that each point of the horizon has the same angular velocity (as measured at the infinity) and in this sense the surface of the black hole is rotating as a rigid body \citep{Frolov:2014dta}.

\section{Geodesics around Rotating Black Holes}\label{Sec3}
\begin{figure*}
\centering
\begin{tabular}{c c}
    \includegraphics[scale=0.52]{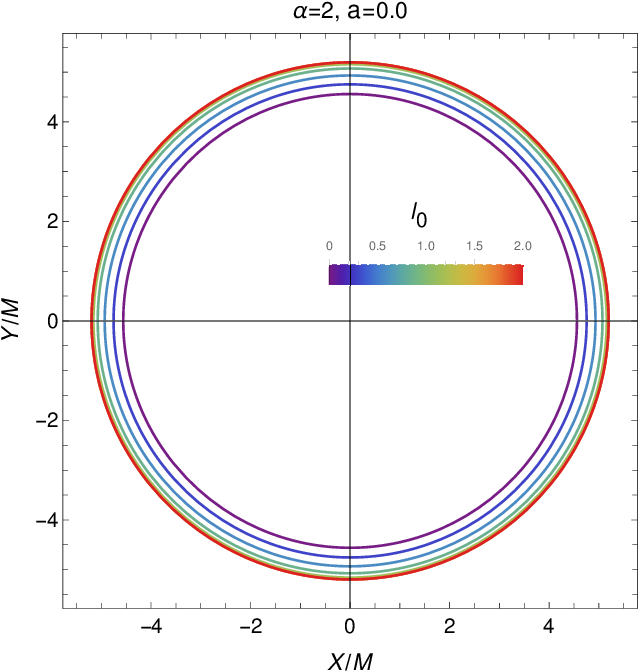}&
     \includegraphics[scale=0.52]{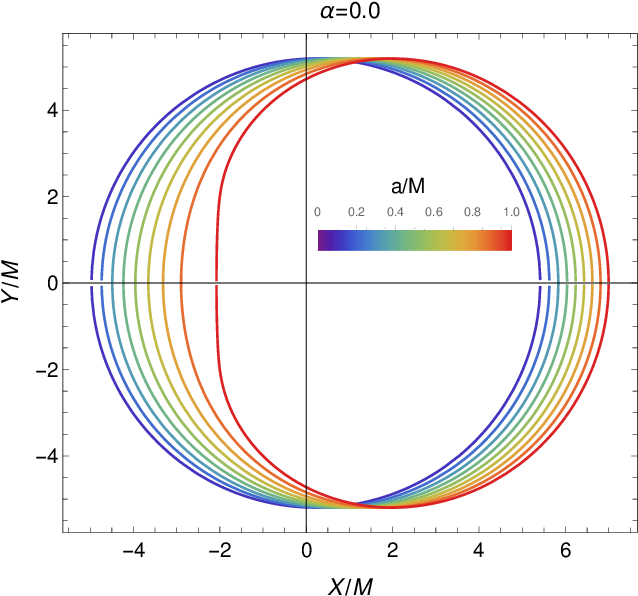}
\end{tabular}
\caption{Shadow geometry of non-rotating hairy black holes with varying $l_0$ parameter (left) and Kerr black hole ($\alpha=0$) with varying spin $a$ (right).}\label{shadow1}
\end{figure*}

The geodesic motion of photons around the black hole is necessary to study the shadow formation, and for this purpose, we consider a motion of test particle in stationary and axially symmetric spacetime. Since metric (\ref{metric}) does not depend on $t$ and $\phi$, hence these are cyclic coordinates and corresponding killing vectors are given by $\chi_{(t)}^{\mu}=\delta _t^{\mu }$  and $\chi_{(\phi)}^{\mu}=\delta _{\phi }^{\mu }$, whose existence further implies that the corresponding four-momentum components $p_{t}$ and $p_{\phi}$ are constants of motion.  The motion of the test particle, neglecting the back reaction, is determined by the rest mass $m_0$, total energy $E$, axial angular momentum $L_z$ and carter constant $\mathcal{Q}$ which is associated with the second-rank irreducible tensor field of hidden symmetry \citep{Carter:1968rr}. We use the Hamilton-Jacobi equation following the integral approach pioneered by Carter \citep{Carter:1968rr} to obtain the geodesic equations in the first-order differential form, which for the metric (\ref{metric}) read \citep{Chandrasekhar:1985kt}
\begin{align}
\Sigma \frac{dt}{d\lambda}=&\frac{r^2+a^2}{\Delta}(E(r^2+a^2)-aL_{z})-a(aE\sin^2{\theta}-L_{z}),\\
\Sigma \frac{d\phi}{d\lambda}=&\frac{a}{\Delta}(E(r^2+a^2)-aL_{z})-(aE-\frac{L_z}{\sin^2{\theta}}),\\
\Sigma \frac{dr}{d\lambda}=&\pm\sqrt{\mathcal{R}(r)}\ ,\label{req} \\
\Sigma \frac{d\theta}{d\lambda}=&\pm\sqrt{\Theta(\theta)}\ ,\label{theq}
\end{align}
where $\lambda$ is the affine parameter along the geodesics and the effective potentials $\mathcal{R} (r)$ and ${\Theta}(\theta)$ for radial and polar motion are given by 
\begin{align}
\mathcal{R}(r)=&E^2\left[\Big((r^2+a^2)-a\xi \Big)^2-\Delta \Big({\eta}+(a-{ \xi})^2\Big)\right],\label{Rpot}\\
\Theta(\theta)=&E^2[\eta-\left(\frac{{ \xi}^2}{\sin^2\theta}-a^2 \right)\cos^2\theta]\ . \label{theta0}
\end{align}
The constant $\mathcal{K}$ is the separability constant related to the Carter constant $\mathcal{Q}$ through $\mathcal{Q}=\mathcal{K}+(aE-L_z)^2$ \citep{Chandrasekhar:1985kt}. We introduce dimensionless quantities called impact parameters \citep{Chandrasekhar:1985kt}
\begin{eqnarray}
    \xi=\frac{L_z}{E} \;\; \text{,} \;\; \eta=\frac{\mathcal{K}}{E^2},
\end{eqnarray}
which are constant along the geodesics. The allowed region around black hole for possible photon motion is $\mathcal{R}\geq 0$ and $\Theta(\theta)\geq 0$ and the sign of $\dot{r}$ and $\dot{\theta}$ can be independently chosen to be either positive or negative. The change of sign occurs at turning points of motion, i.e. when $\mathcal{R}=0$ or $\Theta=0$ \citep{Chandrasekhar:1985kt}.  Depending on the critical parameters' values, the photon may either get captured, scatter to infinity, or form bound orbits. Here, we are interested in spherical lightlike geodesics constrained on a sphere of constant coordinate radius $r$ characterized by $\dot{r}=0$ and $\ddot{r}=0$ called the spherical photon orbits. Mathematically, this corresponds to the local extremum of the radial effective potential outside the horizon at which photons follow unstable orbit at radius $r=r_p$ such that \citep{Frolov:1418196,Chandrasekhar:1985kt}
\begin{eqnarray}
\mathcal{R}=\mathcal{R}'=0 \,\, \text{and}\,\,\mathcal{R}''\leq 0.\label{unstableOrbit} 
\end{eqnarray}
\begin{figure*}
\begin{tabular}{c c}
	\includegraphics[scale=0.52]{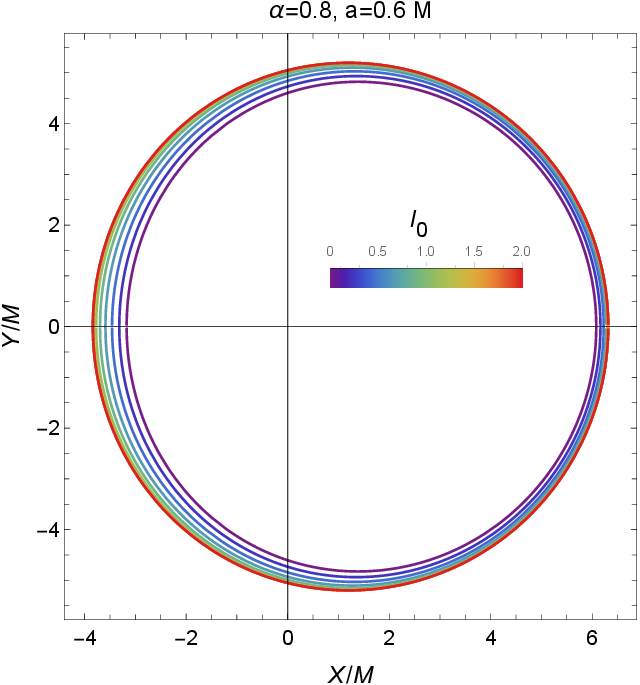}&
	\includegraphics[scale=0.52]{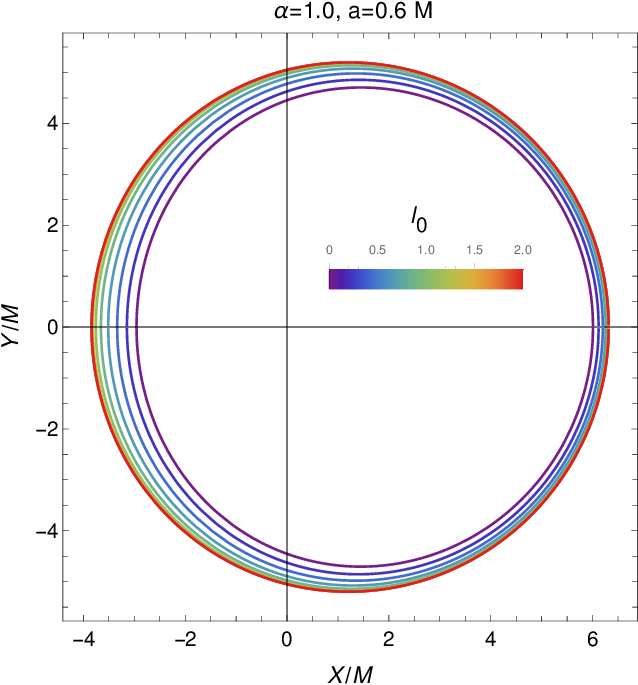}\\
	\includegraphics[scale=0.52]{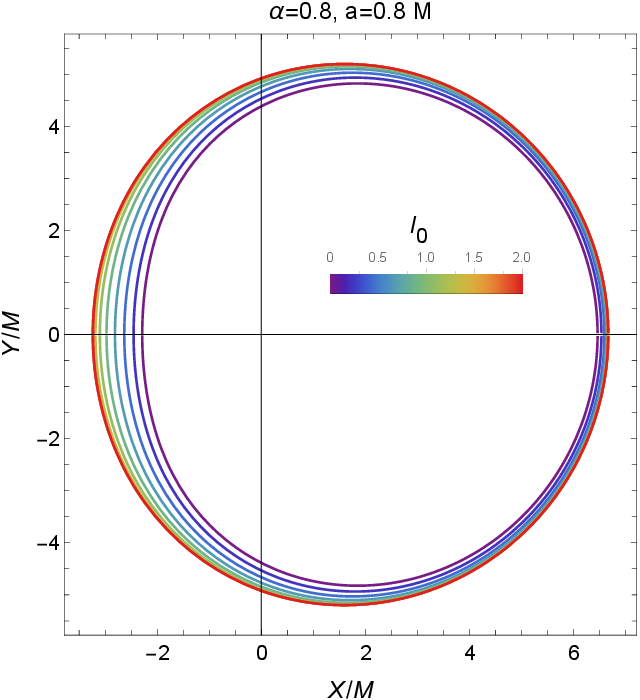}&
	\includegraphics[scale=0.52]{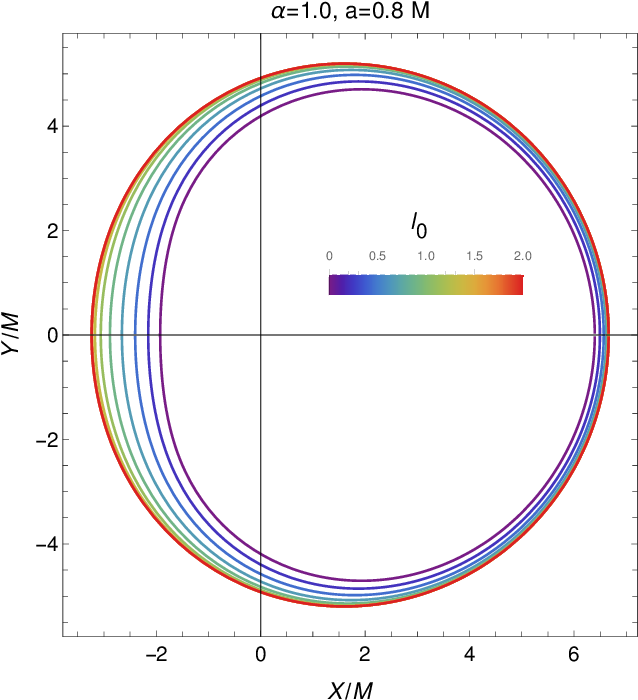}\\
\end{tabular}
\caption{Shadow geometry of hairy Kerr black holes compared with Kerr black hole (outermost solid curve corresponding to $l_0 \to l_k$).}\label{shadow}	
\end{figure*}
\begin{figure*}
\begin{tabular}{c c}
\includegraphics[scale=0.75]{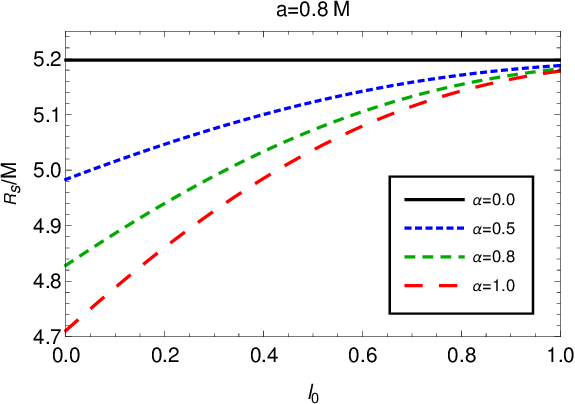}&
\includegraphics[scale=0.77]{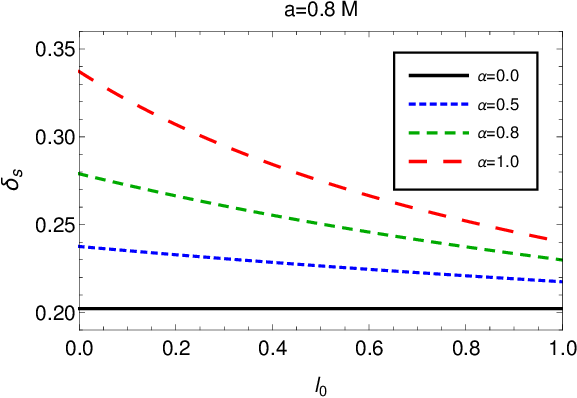}\\
\includegraphics[scale=0.75]{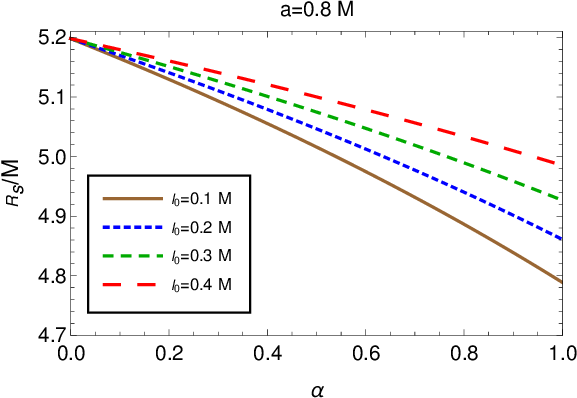}&
\includegraphics[scale=0.77]{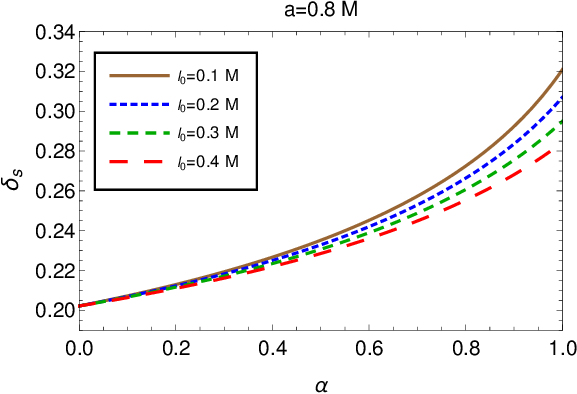}\\
\end{tabular}	
\caption{Variation of the shadow observables $R_s$ (left) and $\delta_s$ (right) of the shadow of hairy Kerr black holes.} \label{Rs_deltaS}
\end{figure*}
Solving equation~(\ref{unstableOrbit})  results into the critical values of impact parameters ($\xi_{crit}, \eta_{crit}$) for the unstable orbits 
\begin{align}
\xi_{crit}=&\frac{\left(a^2+r^2\right) \Delta '(r)-4 r \Delta (r)}{a \Delta '(r)},\nonumber\\
\eta_{crit}=&\frac{r^2 \left(8 \Delta (r) \left(2 a^2+r \Delta '(r)\right)-r^2 \Delta '(r)^2-16 \Delta (r)^2\right)}{a^2 \Delta '(r)^2}\label{CriImpPara},
\end{align}
where $'$ stands for the derivative concerning the radial coordinate, equation~(\ref{CriImpPara}), in the limit $\alpha\to0$, reduces to the critical impact parameter for the Kerr black holes \citep{Hioki:2009na}. The unstable photon orbits have been particularly studied with much interest for black holes and naked singularities \citep{Wilkins:1972rs,Goldstein1974,Johnston:1974pn,Izmailov1979,Izmailov1980,Teo:2020sey} and in the case of the black hole, they define the boundary between the capture and scatter cross-section of light rays. In axially symmetric spacetimes, there can be two circular photon orbits at the equatorial plane viz. those moving in the same direction as the black hole's rotation and those moving opposite to it which are respectively called the prograde and retrograde photons. Owing to the Lens-Thirring effect \citep{Bardeen:1975zz}, whereby, due to the dragging of the inertial frame by the black hole's rotation, to an observer at infinity, the prograde photons must have smaller orbits to account for the excess angular momentum. In contrast to this, the retrograde ones, having lost some angular momentum effectively, would have to rotate at larger radii \citep{Bardeen:1972fi,Teo:2020sey}. The non-planar or 3D photon orbits only arise for $\mathcal{\eta}_{crit}>0$ whereas, for $\mathcal{\eta}_{crit}=0$, the photon orbits are planar and confined to the equatorial plane only. The radii of prograde ($r_{p}^-$) and retrograde ($r_{p}^+$) orbits at the equatorial plane are obtained as roots of $\eta_{crit}=0$. It turns out that these radii for the Kerr black holes are \citep{Teo:2020sey}
\begin{eqnarray}
    r_{p}^-\equiv2M\left[1+\cos\left({\frac{2}{3}\arccos\left(-\frac{|a|}{M}\right)}\right)\right]\ , \\
    r_{p}^+\equiv2M\left[1+\cos\left({\frac{2}{3}\arccos\left(\frac{|a|}{M}\right)}\right)\right]\ ,
\end{eqnarray}
and they fall in the range $M\leq r_p^-\leq 3M$ and  $3M\leq r_p^+\leq 4M$ \citep{Kumar:2018ple}. In the case of Schwarzschild black hole ($a=0$) the two radii degenerate into a photon sphere of constant radius  $r_{p}^-=r_{p}^+=3M$ \citep{Kumar:2020hgm}.  While $r_{p}^+>r_{+}$, all spherical photon orbits are confined to the region $r_{p}^-<r_{p}<r_{p}^+$. For the $\eta_{crit}>0$ scenario, the non-planar ($\theta\neq\frac{\pi}{2}$ and $\dot{\theta}\neq0$) geodesics have an additional latitudinal motion governed by the Carter constant which accounts for a hidden spacetime symmetry. Furthermore, these orbits oscillate symmetrically about the equatorial plane while repeatedly crossing it \citep{Teo:2020sey}. 

\section{Shadow of hairy Kerr black holes}\label{Sec4}
The black hole shadow is the optical appearance of the black hole caused by the strong gravitational lensing \citep{Virbhadra:1999nm,Bozza:2001xd,Bozza:2002zj,Ghosh:2020spb,Kumar:2020sag,Islam:2020xmy} of light near these compact objects and appears as a 2D dark zone for a far distant observer (for details see \citep{Grenzebach:2014fha,Cunha:2018acu,Cunha:2018gql,Huang:2018rfn} and references therein). The shape and the size of the shadows depend on the parameters associated with the black hole spacetime geometry. The shadow, in turn, has facilitated the estimation and measurement of various black hole parameters like its mass, spin angular momentum as well as other \textit{hairs} \citep{Kumar:2018ple}. Thus, it is a tool to test Einstein's GR in the strong-field regime \citep{Gott:2018ocn,Kumar:2020owy}, and also the no hair theorem \citep{Cunha:2015yba}. The photon region around the black hole's event horizon is formed from the combination of all unstable spherical photon orbits, i.e. the separatrix between photon geodesics that escape to spatial infinity and those that fall into the event horizon. 
We assume uniformly distributed light sources at infinity and the photons coming with all possible impact parameters either get scattered near the vicinity of the black hole (\ref{metric}) or captured by it. We also assume a distant observer at an inclination angle $\theta_0$ with the rotation axis of the black hole. The celestial coordinates ($X$,$Y$) of the shadow boundary at the observer's sky are the apparent angular distances of the image measured from the line of sight in directions perpendicular and parallel respectively, to the projected axis of rotation black hole onto the celestial sphere \citep{Hioki:2009na}. By making a stereographic projection of black hole shadow on the observer's celestial sky to the image plane, the shadow boundary can be described by the following coordinates
\begin{align}
X=&\lim_{r_o\rightarrow\infty}\left(-r_o^2 \sin{\theta_o}\frac{d\phi}{d{r}}\right),\nonumber\\ Y=&\lim_{r_o\rightarrow\infty}\left(r_o^2\frac{d\theta}{dr}\right),\label{Celestial1}
\end{align} 
where $r_0$ is the distance between the observer and the black hole. For an asymptotically far observer, equation (\ref{Celestial1}) leads to
\begin{align}
X=&-\xi_{crit}\csc\theta_o,\nonumber\\
Y=&\pm\sqrt{\eta_{crit}+a^2\cos^2\theta_o-\xi_{crit}^2\cot^2\theta_o}\ .\label{Celestial2}    
\end{align}
Further, if the observer is at the equatorial plane ($\theta_o=\pi/2$), it simplifies to
\begin{eqnarray}
X=-\xi_{crit}\ , \nonumber\\ 
Y=\pm\sqrt{\eta_{crit}}\ .\label{Celestial3}
\end{eqnarray}
The contour of $X$ and $Y$ delineates the shadow for the hairy Kerr black holes. 
Equation~(\ref{Celestial3}), on using metric (\ref{metric1}), yields
\begin{eqnarray}
X=&-&\frac{[a^2 - 3 r_p^2] m(r_p) + r_p [a^2 + r_p^2] [1 + m'(r_p)]}{a [m(r_p) + r_p [-1 + m'(r_p)]]}, \nonumber\\
Y=&\pm&\frac{1}{a [m(r_p) + r_p [-1 + m'(r_p)]]}\Big[r_p^{3/2}\Big[-r_p^3(1+m'(r_p)^2)\nonumber\\
 &+&m(r_p) [4 a^2 + 6 r_p^2 - 9 r_pm(r_p)] - 2r_p [2 a^2 + r_p^2 \nonumber\\
 &-&3 r_pm(r_p)] m'(r_p)\Big]^{1/2}\Big]\label{Celestial4},
\end{eqnarray}
where for brevity, we have used $m(r)=M-\frac{\alpha}{2}r e^{-r/(M-\frac{l_0}{2})}$. Whereas for the Kerr case, equation~(\ref{Celestial4}) simplifies to 
\begin{eqnarray}
X &=&\frac{r_p^2 (r_p-3 M ) + a^2 (M + r_p)}{a (r_p-M)},\nonumber\\
Y &=&\pm\frac{r_p^{3/2} (4 a^2 M - r_p (r_p-3 M )^2)^{1/2}}{a( r_p-M)}
\end{eqnarray}
which is exactly the same as that obtained for the Kerr black hole \citep{Hioki:2009na}, which for the  nonrotating black hole  (\ref{metric1}) takes the form
\begin{equation}\label{nr}
X^2+Y^2=\frac{2 r_p^4 + [m(r_p) + r_p m'(r_p)][-6 r_p^2 m(r_p) + 2 r_p^3 m'(r_p)]}{[m(r_p) + r_p[-1 + m'(r_p)]]^2}. 
\end{equation}
The equation~(\ref{nr}) ensures that the shadow of a nonrotating black hole (\ref{metric1}) is perfectly circular and further, equation~(\ref{nr}) reverts to $X^2+ Y^2=27M^2$ for the Schwarzschild black hole.
\begin{figure*}
	\begin{center}
		\begin{tabular}{c c}
			\includegraphics[scale=0.7]{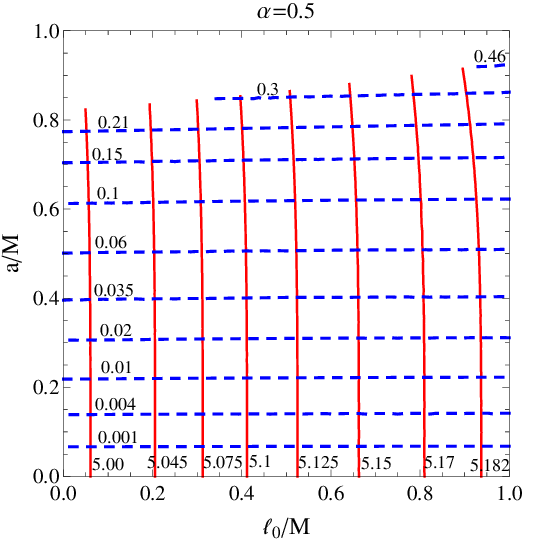}&
			\includegraphics[scale=0.7]{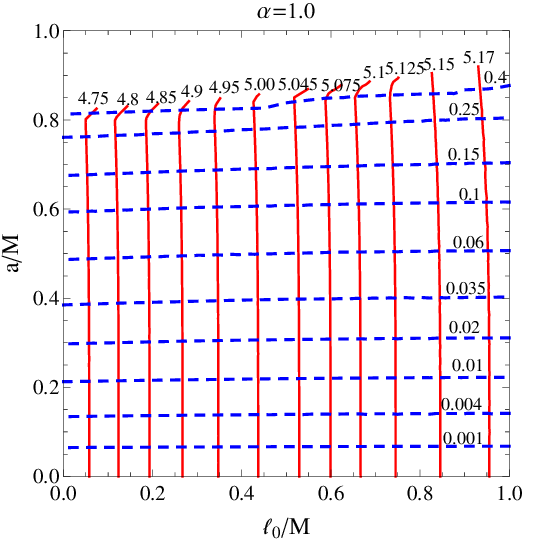}
		\end{tabular}	
	\end{center}
    \caption{The contour maps for hairy Kerr black hole shadow observables $R_s$ (red solid lines) and $\delta_s$ (blue dashed lines) at inclination angle $\theta_0=90\si{\degree}$.}
    \label{Rs_ds_Estimation}
\end{figure*}
\begin{table*}
    \caption{Estimated values of hairy Kerr black hole parameters $a/M$ and $l_0/M$ from known shadow observables $R_s$ and $\delta_s$ for different values of $\alpha$ at inclination angle $\theta=90\si{\degree}$.}
    \begin{subtable}{.5\linewidth}
      \centering
        \caption{$\alpha=0.5$}
       \begin{tabular}{||c c c c||} 
 \hline
 $R_s$ & $\delta_s$ & $a/M$ & $l_0/M$ \\ [0.8ex] 
 \hline\hline
 \;\; 5.000\;\; &\;\; 0.001\;\; &\;\; 0.0682\;\; &\;\; 0.0608\;\; \\ [2ex] 
 \hline
 \;\; 5.075\;\; &\;\; 0.010\;\; &\;\; 0.2225\;\; &\;\; 0.3119\;\; \\ [2ex] 
 \hline
 \;\; 5.125\;\; &\;\; 0.035\;\; &\;\; 0.4031\;\; &\;\; 0.5235\;\; \\ [2ex] 
 \hline
 \;\; 5.150\;\; &\;\; 0.100\;\; &\;\; 0.6204\;\; &\;\; 0.6557\;\; \\ [2ex] 
 \hline
 \;\; 5.182\;\; &\;\; 0.210\;\; &\;\; 0.7923\;\; &\;\; 0.9113\;\; \\ [2ex] 
 \hline
\end{tabular}
    \end{subtable}%
    \begin{subtable}{.5\linewidth}
      \centering
        \caption{$\alpha=1.0$}
        \begin{tabular}{||c c c c||} 
 \hline
 $R_s$ & $\delta_s$ & $a/M$ & $l_0/M$ \\ [0.8ex] 
 \hline\hline 
 \;\; 4.750\;\; &\;\; 0.001\;\; &\;\; 0.0657\;\; &\;\; 0.0583\;\; \\ [2ex] 
 \hline
 \;\; 4.900\;\; &\;\; 0.010\;\; &\;\; 0.2181\;\; &\;\; 0.2669\;\; \\ [2ex] 
 \hline
 \;\; 5.045\;\; &\;\; 0.035\;\; &\;\; 0.3986\;\; &\;\; 0.5289\;\; \\ [2ex] 
 \hline
 \;\; 5.125\;\; &\;\; 0.100\;\; &\;\; 0.6147\;\; &\;\; 0.7405\;\; \\ [2ex] 
 \hline
 \;\; 5.170\;\; &\;\; 0.250\;\; &\;\; 0.8056\;\; &\;\; 0.9387\;\; \\ [2ex] 
 \hline
\end{tabular}
    \end{subtable} 
\label{parameter_table1}
\end{table*}
\begin{figure*}
\begin{tabular}{c c}
	\includegraphics[scale=0.55]{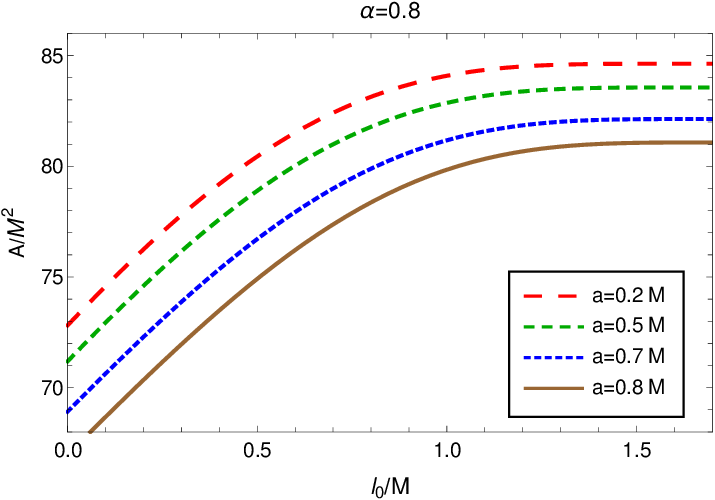}&
	\includegraphics[scale=0.55]{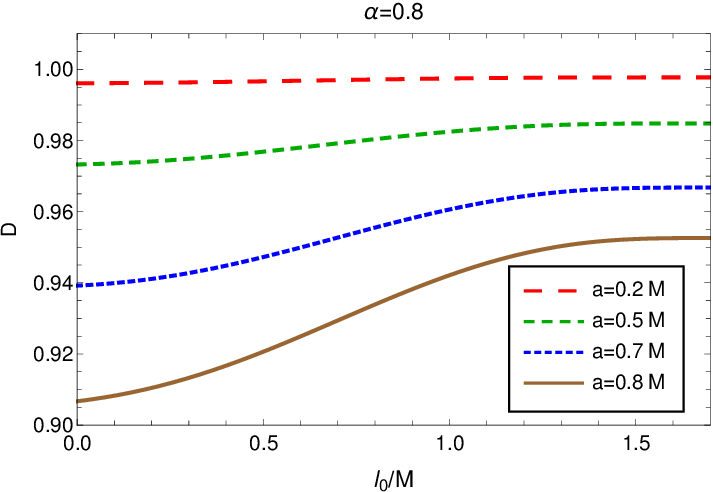}\\
	\includegraphics[scale=0.55]{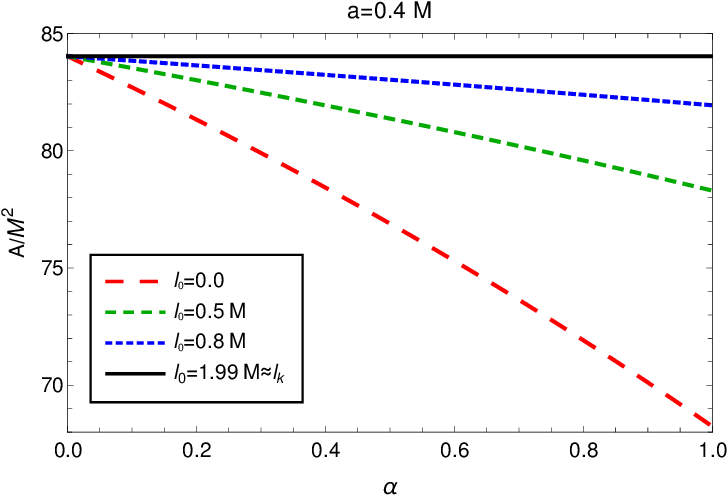}&
	\includegraphics[scale=0.55]{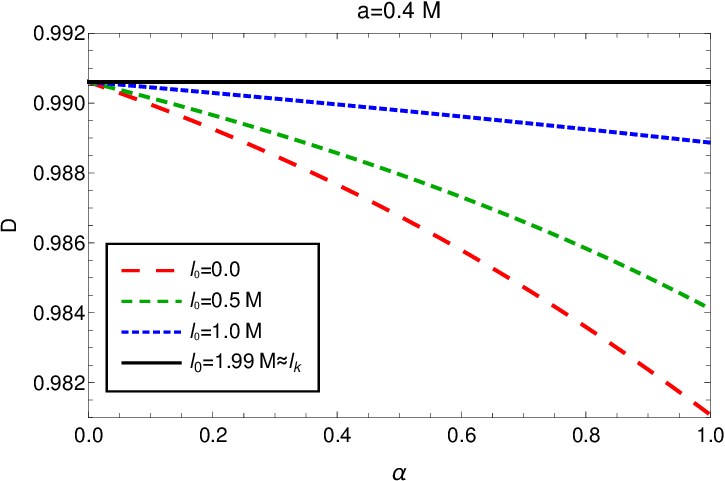}
\end{tabular}
\caption{The shadow area $A$ and oblateness $D$ with varying parameters for the hairy Kerr black holes. The black solid line in lower panels correspond to Kerr black hole.}\label{obs}
\end{figure*}

The shadows for the non-rotating case ($a=0$) and Kerr black holes ($\alpha=0$) are depicted in Fig.~\ref{shadow1}. Whereas, a comparison of the hairy Kerr black hole shadows with that of the Kerr black hole is shown in Fig.~\ref{shadow}. The decrease in $l_0$ from the Kerr limit $l_0=l_k$, causes a decrease in the shadow size from that of the Kerr black hole (shown by the outermost solid circles in Fig.~\ref{shadow}). Besides, the parameter $\alpha$ is found to have a decremental effect on the size of the hairy Kerr black hole shadow size, as for fixed values of $a$ and $l_0$, the shadow size decreases with the increase in $\alpha$.

\section{Observables and black hole parameter estimation}\label{Sec5}
The parameters associated with the hairy Kerr black holes are expected to be constrained from \textit{EHT} observations. The observed image of supermassive black hole M87* is as per the Kerr black hole of GR. However, the \textit{EHT} observation did not mention black holes in MoG \citep{Akiyama:2019cqa,Akiyama:2019bqs,Akiyama:2019eap}, which is what we intend to probe with the estimated parameters of our hairy Kerr black holes. We assume that the observer is in the equatorial plane, i.e., at an inclination angle $\theta_0=\pi/2$ for the parameter estimation. 
\paragraph{Hioki and Maeda method:}
\cite{Hioki:2009na} characterized the black hole shadow distortion and size by proposing the two observables, viz. $R_s$ and $\delta_s$. They approximate the shadow by a reference circle,  with $R_s$ being the radius of this circle, and $\delta_s$, the deviation of the left edge of the shadow from the circle boundary \citep{Hioki:2009na}. 
The shadow reference circle  coincides at the top ($X_t$, $Y_t$), bottom ($X_b$, $Y_b$) and right ($X_r, 0$) edges with the shadow contour \citep{Ghosh:2020ece},
where ($X^{'}_l$, 0) and ($X_l$, 0) are respectively the points where the reference circle and leftmost edge of the shadow contour intersects the horizontal axis. Indeed, while $R_s$ designates approximately the size of the shadow, $\delta_s$ connotes the shadow deformation from the reference circle. These observables are defined as \citep{Hioki:2009na}
\begin{eqnarray}
    R_s=\frac{(X_t-X_r)^2+Y_{t}^2}{2|X_r-X_t|},
\end{eqnarray}\label{Rs}
using the relations $X_b=X_t$ and $Y_b=-Y_t$, and
\begin{eqnarray}
    \delta_s=\frac{|X_l-X^{'}_l|}{R_s}\ ,
\end{eqnarray}\label{deltas}
where subscripts $r$, $l$, $t$ and $b$, respectively, stand for the right, left, top and bottom of the shadow boundary.

The presence of parameters $\alpha$ and $l_0$ has a profound influence on the approximate size and shape of the shadow (cf. Fig.~\ref{Rs_deltaS}). The shadow radius $R_s$ of hairy Kerr black holes increases monotonically with the increase in $l_0$, but interestingly $\alpha$ has a decremental effect, and the shadow radius of hairy Kerr black holes are always smaller than that of the Kerr black hole (solid black curve in Fig.~\ref{Rs_deltaS}), which is also evident from the shadow templates shown in Fig.~\ref{shadow}. In contrast, the shadow distortion parameter $\delta_s$ of hairy Kerr black holes decreases with $l_0$ but increases with $\alpha$, i.e., the shadows of hairy Kerr black holes are more distorted than the Kerr black hole (solid black curve in Fig.~\ref{Rs_deltaS}), and this trend is again inferable from Fig.~\ref{shadow}. 

To estimate the black hole parameters, we plot the contours of $R_s$ and $\delta_s$ in the ($l_0-a$) parameter space of the hairy Kerr black hole in Fig.~\ref{Rs_ds_Estimation}. It is evident from the Fig.~\ref{Rs_ds_Estimation}, the curves of radius $R_s$ and distortion $\delta_s$ intersect each other at a particular point, which for a given $\alpha$ depends on both $a$ and $l_0$. An intersection point of $R_s$ and $\delta_s$ for a given $\alpha$ and inclination angle $\theta_0$, uniquely determines the parameters $a$ and $l_0$ (cf. Table~\ref{parameter_table1}). Thus, we have extended the \cite{Hioki:2009na} method to unambiguously determine the parameters $a$ and $l_0$ of the hairy Kerr black holes. 

\paragraph{Kumar and Ghosh method:}
The Hioki and Maeda method \citep{Hioki:2009na} of numerical estimation of black hole parameters using shadow observables $R_s$ and $\delta_s$ was later extended by \cite{Tsupko:2017rdo} to analytical estimation. Further, \cite{Tsukamoto:2014tja} proposed a method to distinguish the Kerr black hole shadow from the other rotating black hole shadows arising in the MoG. However, the observables $R_s$ and $\delta_s$ used in the above two methods demand some symmetries in the shadow shape, and they may not work precisely for the haphazard black hole shadows in MoG \citep{Schee:2008kz,Johannsen:2015qca,Tsukamoto:2014tja, Abdujabbarov:2015xqa,Younsi:2016azx, Tsupko:2017rdo}. Also, the shadow shape may not be precisely circular due to noisy data \citep{Abdujabbarov:2015xqa,Kumar:2018ple}. Motivated by this, \cite{Kumar:2018ple} presented observables characterizing a haphazard (or not circular shape) shadow to estimate the parameters associated with black holes. They defined the shadow Area ($A$) and oblateness ($D$) by
\begin{eqnarray}
A=2\int{Y(r_p) dX(r_p)}=2\int_{r_p^{-}}^{r_p^+}\left( Y(r_p) \frac{dX(r_p)}{dr_p}\right)dr_p,\label{Area}
\end{eqnarray} 
and
\begin{eqnarray}
D=\frac{X_r-X_l}{Y_t-Y_b}.\label{Oblateness}
\end{eqnarray}
For an equatorial observer, the oblateness $D$ may vary between $\sqrt{3}/2\leq D<1$, with $D=1$ for the Schwarzschild case ($a=0$) and $D=\sqrt{3}/2$ for extremal Kerr case ($a=M$) \citep{Tsupko:2017rdo}. 

We plot the observables $A$ and $D$ in Fig.~\ref{obs}, and it is evident that all the three parameters $a$, $\alpha$ and $l_0$ have a profound impact on the shadow area. Both $A$ and $D$ increase with $l_0$ for  fixed values of $\alpha$ and $a$ and decreases with $\alpha$ for fixed values of $l_0$ and $a$. Both $A$ and $D$ decrease with $a$ for fixed values of $\alpha$ and $l_0$. Moreover, evidently from the lower panels in Fig.~\ref{obs}, the shadow area of the hairy Kerr black holes is smaller than that of the Kerr black hole and also is more distorted, which can also be checked from Fig.~\ref{shadow}.

\begin{figure*}
		\begin{tabular}{c c}
			\includegraphics[scale=0.7]{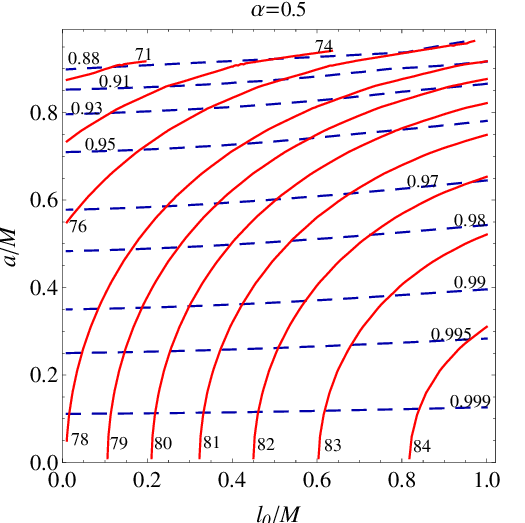}&
			\includegraphics[scale=0.7]{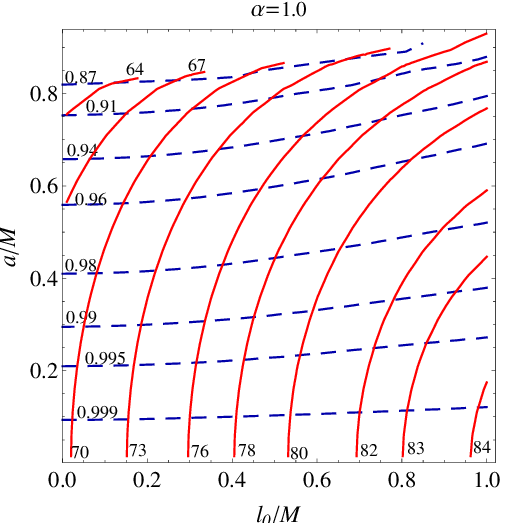}
		\end{tabular}
	\caption{Contour plots of the observables $A/M^2$ and $D$ in the hairy Kerr black hole parameter space ($a/M,l_0/M$) for different $\alpha$. Solid red curves correspond to constant $A$, and dashed blue curves are for constant oblateness parameter $D$. }
	\label{parameterestimation}
\end{figure*}

Furthermore, we plot the contour map of the observables $A$ and $D$ in the ( $l_0$-$a$) plane (Fig.~\ref{parameterestimation}). It is clear from the Fig.~\ref{parameterestimation} that the intersection points give exact values of parameters $a$ and $l_0$ for the chosen value of $A$ and $D$. In the Table~\ref{parameter_table2} we have shown the estimated values of parameters $a$ and $l_0$ of hairy Kerr black holes for given shadow observables $A$ and $D$.
\begin{table*}
    \caption{Estimated values of hairy Kerr black hole parameters $a/M$ and $l_0/M$ from known shadow observables $A$ and $D$ for different values of $\alpha$.}
    \begin{subtable}{.5\linewidth}
      \centering
        \caption{$\alpha=0.5$}
       \begin{tabular}{||c c c c||} 
 \hline
 $A/M^2$ & $D$ & $a/M$ & $l_0/M$ \\ [0.8ex] 
 \hline\hline
 \;\; 84\;\; &\;\; 0.999\;\; &\;\; 0.1250\;\; &\;\; 0.8815\;\; \\ [2ex] 
 \hline
 \;\; 83\;\; &\;\; 0.995\;\; &\;\; 0.2714\;\; &\;\; 0.7172\;\; \\ [2ex] 
 \hline
 \;\; 82\;\; &\;\; 0.980\;\; &\;\; 0.5233\;\; &\;\; 0.7616\;\; \\ [2ex] 
 \hline
 \;\; 80\;\; &\;\; 0.950\;\; &\;\; 0.7644\;\; &\;\; 0.8038\;\; \\ [2ex] 
 \hline
 \;\; 78\;\; &\;\; 0.910\;\; &\;\; 0.9108\;\; &\;\; 0.9193\;\; \\ [2ex] 
 \hline
\end{tabular}
    \end{subtable}%
    \begin{subtable}{.5\linewidth}
      \centering
        \caption{$\alpha=1.0$}
        \begin{tabular}{||c c c c||} 
 \hline
  $A/M^2$ & $D$ & $a/M$ & $l_0/M$ \\ [0.8ex] 
 \hline\hline
 \;\; 84\;\; &\;\; 0.999\;\; &\;\; 0.1211\;\; &\;\; 0.9790\;\; \\ [2ex] 
 \hline
 \;\; 83\;\; &\;\; 0.990\;\; &\;\; 0.3706\;\; &\;\; 0.9282\;\; \\ [2ex] 
 \hline
 \;\; 80\;\; &\;\; 0.960\;\; &\;\; 0.6587\;\; &\;\; 0.8244\;\; \\ [2ex] 
 \hline
 \;\; 78\;\; &\;\; 0.940\;\; &\;\; 0.7439\;\; &\;\; 0.7515\;\; \\ [2ex] 
 \hline
 \;\; 67\;\; &\;\; 0.870\;\; &\;\; 0.8291\;\; &\;\; 0.2678\;\; \\ [2ex] 
 \hline
\end{tabular}
    \end{subtable} 
\label{parameter_table2}
\end{table*}

\subsection{Energy Emission}
\begin{figure*}
		\begin{tabular}{c c}
			\hspace{-4mm}\includegraphics[scale=0.8]{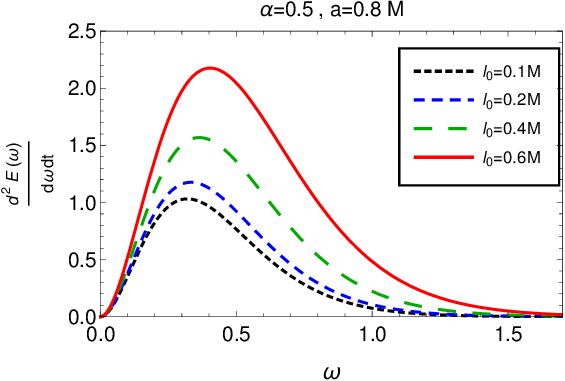}&
			\includegraphics[scale=0.8]{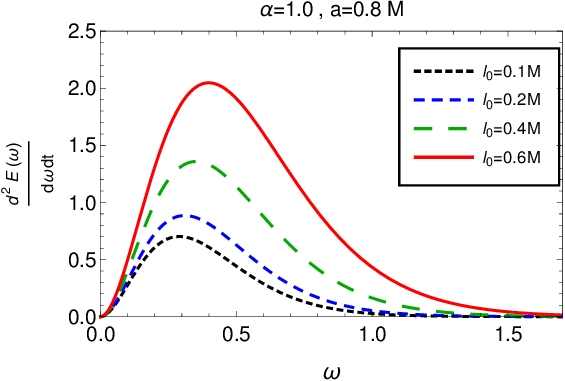}
		\end{tabular}	
	\caption{Evolution of the emission rate with the frequency $\omega$ for different values of parameter $l_0$ and $\alpha$ for the hairy Kerr black holes.}
	\label{EnergyEmission}
\end{figure*}

Observing the shadow can leads to witnessing exciting phenomena, and in the present study, we turn to analyse the energy emission rate for hairy Kerr black holes. For an observer at infinity, the black hole shadow corresponds to a high energy absorption cross-section, which oscillates around a constant limiting value $\sigma_{lim}$ for a spherically symmetric black hole. For a black hole with a photon sphere, the $\sigma_{lim}$ is the same as the geometrical cross-section of the photon sphere \citep{Wei:2013kza} taking the form
\begin{eqnarray}
    \sigma_{lim}\approx\pi R_{s}^2.
\end{eqnarray}
We intend to discuss the energy emission rate of hairy Kerr black holes
(\ref{metric}) using the relation \citep{Wei:2013kza} 
\begin{eqnarray}
\frac{d^2E(\omega)}{d\omega dt}=\frac{2\pi^2R_{s}^2}{e^{\omega/T}-1}\omega^3,
\end{eqnarray}\label{emission_eq} 
where $\omega$ is photon frequency and $T$ is the Hawking temperature at event horizon $r_{+}$ of black hole (\ref{metric}) given by
\begin{eqnarray}
    T=\frac{1}{2\pi \left(a^2+r_{+}^2\right)} \left(\frac{\alpha e^{\frac{2r_{+}}{l_0-2 M}} r_{+}(l_0-2 M+r_{+})}{l_0-2 M}+r_{+}-M\right).
\end{eqnarray}
Interestingly, the Hawking temperature of hairy Kerr black holes depends on the parameter $\alpha$ as well as on $l_0$. In Fig.~\ref{EnergyEmission}, we demonstrate the behaviour of the energy emission rate of hairy Kerr black holes against the photon frequency $\omega$ for various values of parameter $l_0$ for a given $\alpha$ and $a$. The emission rate peak decreases with an increase in $l_0$ and shifts to a lower value of $\omega$ while an increase. 
\begin{figure*}
	\begin{center}
		\begin{tabular}{c c}
			\includegraphics[scale=0.65]{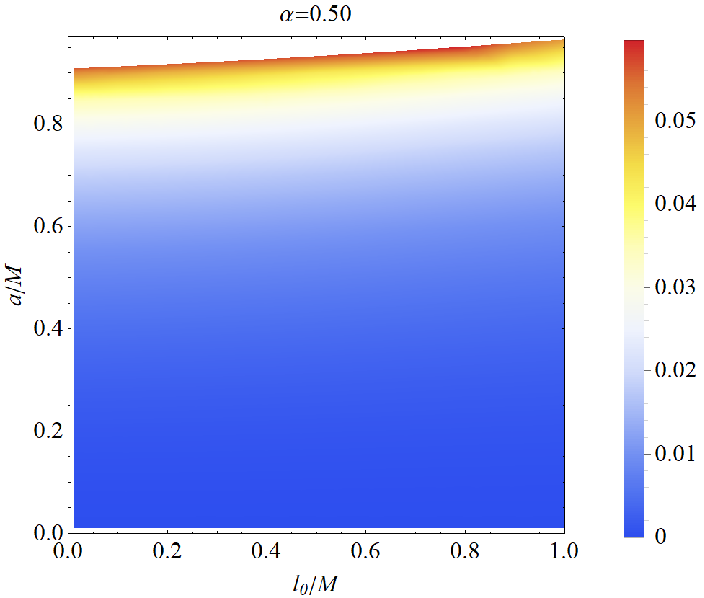}&
			\includegraphics[scale=0.65]{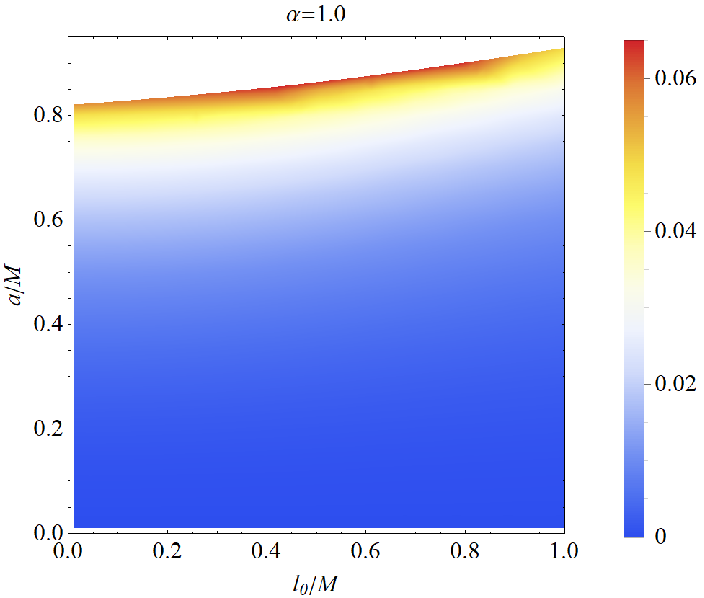}\\
			\includegraphics[scale=0.65]{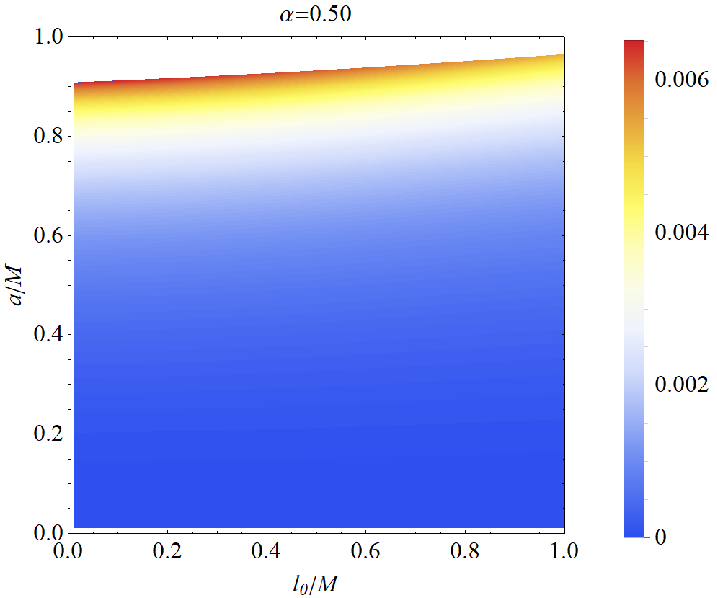}&
			\includegraphics[scale=0.65]{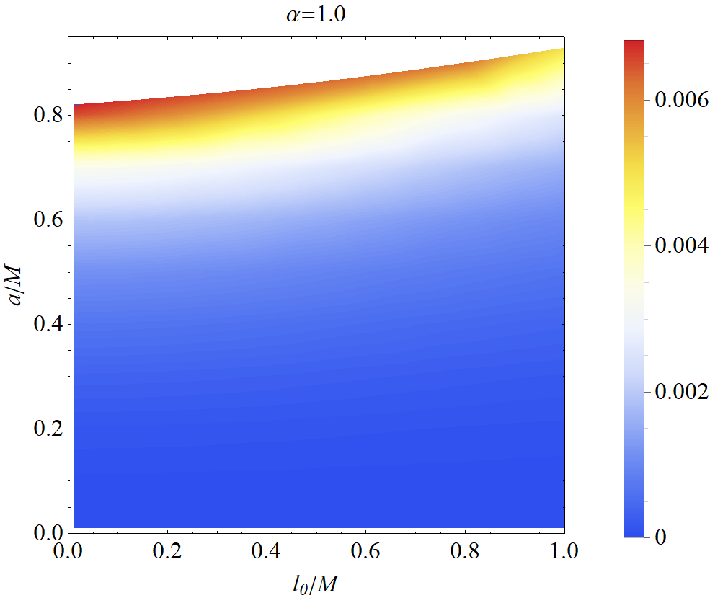}\\
		\end{tabular}	
	\end{center}
	\caption{Circularity deviation observable $\Delta C$ for hairy Kerr black hole shadows as a function of parameters ($a/M$ and $l_0/M$) in agreement with the \textit{EHT} observations of the M87* black hole, $\Delta C \leq 0.1$, is satisfied for the entire parameter space ($a/M$ and $l_{0}/M$). The parameters of M87* used are $M=6.5\times 10^9 M_{\odot}$ and $d=16.8$MPc. The inclination angle is $\theta_0=90\si{\degree}$ (upper panel) and $17\si{\degree}$ (lower panel). The white region is forbidden for ($a/M$ and $l_0/M$).}
	\label{M87obs1}
\end{figure*}
\begin{figure*}
	\begin{center}
		\begin{tabular}{c c}
			\includegraphics[scale=0.62]{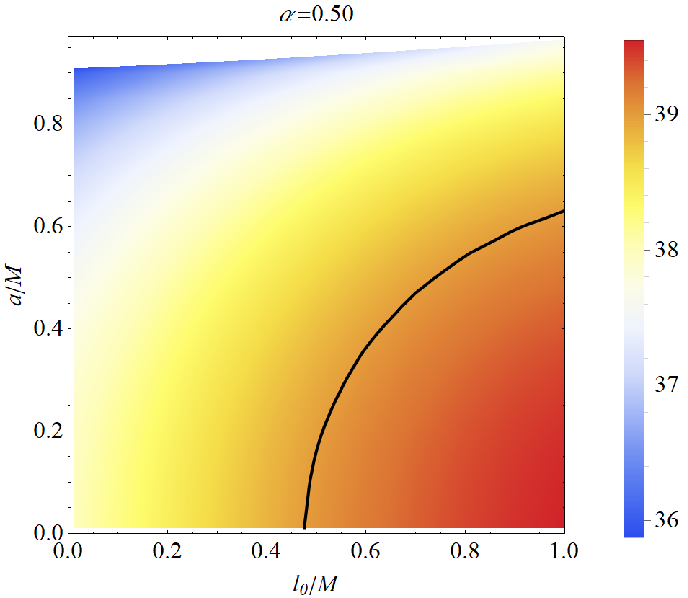}&
			\includegraphics[scale=0.62]{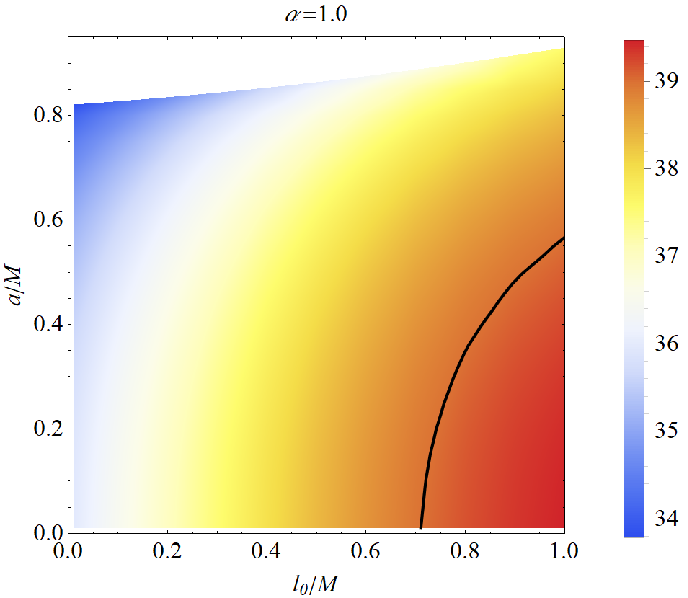}\\
			\includegraphics[scale=0.62]{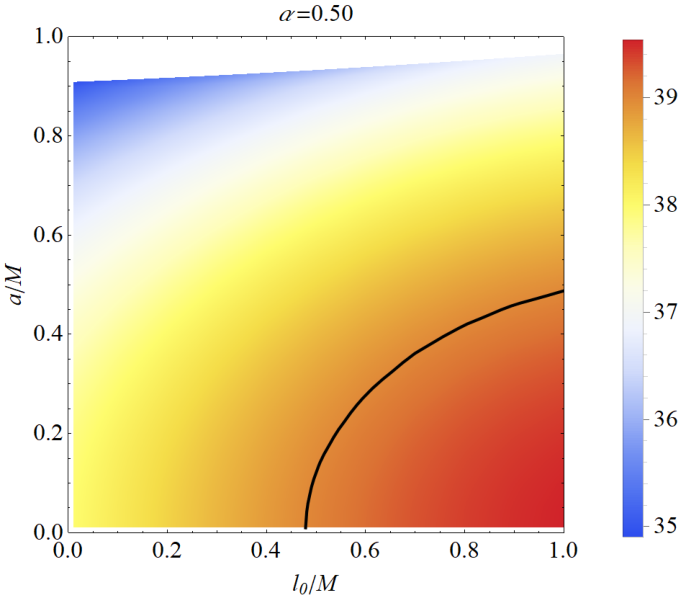}&
			\includegraphics[scale=0.62]{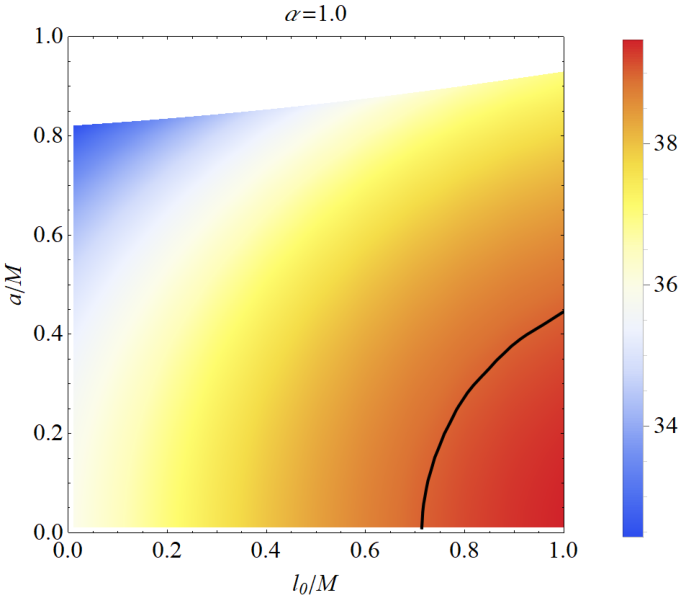}
		\end{tabular}	
	\end{center}
	\caption{Angular diameter observable $\theta_d$ for hairy Kerr black hole shadows as a function of parameters ($a/M$ and $l_0/M$). The black solid curve correspond to $\theta_d=39\mu$as within $1 \sigma$ region of the measured angular diameter, $\theta_d = 42 \pm 3 \mu$as, of the M87* black hole reported by the \textit{EHT}. The parameters of M87* used are $M=6.5\times 10^9 M_{\odot}$ and $d=16.8$Mpc. The inclination angle is $\theta_0=90\si{\degree}$ (upper panel), $17\si{\degree}$ (lower panel). The white region is forbidden for ($a/M$ and $l_0/M$).}
	\label{M87obs2}
\end{figure*}
\section{Constraints from \textit{EHT} observation of M87*}\label{Sec6}
The \textit{EHT} collaboration using the VLBI technology unveiled the event horizon scale image of the supermassive black hole M87* \citep{Akiyama:2019cqa,Akiyama:2019bqs,Akiyama:2019eap}. This provides an exciting arena to investigate gravity in the strong-field regime to determine in turn the exact nature of black holes.
 The observed image of M87* is nearly circular and of crescent shape; the circularity deviation is $\Delta C\leq0.10$ (10\%) in terms of root-mean-square deviation from average shadow radius and the angular size $\theta_d$ of the emission region in the observed image is $42 \pm 3\mu as$ \citep{Akiyama:2019cqa,Akiyama:2019bqs,Akiyama:2019eap}. The observed image is constrained with the modelled image of Kerr black hole \citep{Akiyama:2019cqa,Akiyama:2019bqs,Akiyama:2019eap}. It can be further used to constrain the models of black holes in MoG.

The boundary of a black hole shadow can be defined by the polar coordinates ($R(\varphi), \varphi$), with origin at the shadow centre ($X_c$,$Y_c$) such that $X_c=(X_r-X_l)/2$ and $Y_c=0$. The average shadow radius $\bar{R}$ is given by \citep{Bambi:2019tjh} 
\begin{eqnarray}
\bar{R}^2=\frac{1}{2\pi}\int_{0}^{2\pi} R^2(\varphi) d\varphi,
\end{eqnarray}
where 
\begin{eqnarray}
R(\varphi)=\sqrt{(X-X_{c})^2+(Y-Y_{c})^2}\ \text{and} \;\ \varphi\equiv \tan^{-1}\left(\frac{Y}{X-X_C}\right)\nonumber.
\end{eqnarray}
Here,  $\varphi$ is  the angle between the x-axis and the vector connecting the centre $(X_c,Y_c)$ with the point $(X, Y)$ at the boundary of the shadow.
The circularity deviation $\Delta C$ is defined as \citep{Bambi:2019tjh}
\begin{eqnarray}\label{circularity}
\Delta C=\frac{1}{\bar{R}}\sqrt{\frac{1}{2\pi}\int_0^{2\pi}\left(R(\varphi)-\bar{R}\right)^2d\varphi},
\end{eqnarray}
The $\Delta C$ measures deviation from a perfect circle. It will be useful to perform a comparison between the theoretical predictions for hairy Kerr BH  shadows, and the \textit{EHT} observations \citep{Bambi:2019tjh}. 

The angular diameter of the shadow $\theta_d$ \citep{Kumar:2020owy} is defined as 
\begin{eqnarray}
\theta_d=2\frac{R_a}{d} \;\;,\;\; R_a=\sqrt{A/\pi},
\end{eqnarray}  

Assuming M87* a hairy Kerr black hole, we can calculate $\Delta C$ for metric (\ref{metric})  and use \textit{EHT} observation results to put constraints on the hairy Kerr black hole parameters. We shall use the deviation from circularity defined above in equation~(\ref{circularity}) to perform a comparison between predicted hairy Kerr black hole shadow and the \textit{EHT} observations. Assuming Kerr black hole geometry of the M87*, the \textit{EHT} observations demonstrated that $\Delta C \leq 0.1$. Here we shall use hairy Kerr black hole instead. The equation~(\ref{circularity}) implies that the $\Delta C$ depends on the black hole metric parameters.

Relativistic jets, which possibly emerge from magnetohydrodynamic interactions between accretion disks and rotating black holes, are observed in the M87* image. Considering the orientation of these jets in M87*, the inclination angle (with respect to the line of sight) is estimated to be $17\si{\degree}$ \citep{Walker:2018muw}. 

In order to proceed further, a precise estimate of the mass of the supermassive black hole M87*, in question, is required. Ideally, independent determinations of the supermassive black hole mass should agree with each other. Unfortunately, this issue is far from being settle even with current \textit{EHT} supermassive black hole mass determinations \citep{Vagnozzi:2020quf}. However, using a mass of M87*, $M=6.5\times 10^9 M_{\odot}$ as estimated by \textit{EHT} observations \citep{Akiyama:2019cqa,Akiyama:2019eap} and the distance from the earth, $d=16.8$ Mpc, the circularity deviation $\Delta C$ is depicted for inclination angles $90\si{\degree}$ (upper panel) and $17\si{\degree}$ (lower panel) in Fig.~\ref{M87obs1}. Clearly, the Fig.~\ref{M87obs1} indicates that the hairy Kerr black hole shadows of M87* for $\alpha= 0.50, 1.0$, satisfy $\Delta C \leq 0.10$ for entire parameter space ($a/M$, $l_0/M$) at $\theta_0=90\si{\degree}$ and $\theta_0=17\si{\degree}$.

Next we calculate the angular diameter of the shadow, which apart from $a$, $l_0$, $\alpha$, $\theta$, depends on mass $M$ and the distance $d$ of the black hole. The mass of the black hole is $M\sim 6.5 \times 10^9 M_\odot$ using GR Kerr model. The angular diameter $\theta_d=2 R_a/d$, with $d=16.8$Mpc for the hairy Kerr black hole shadows is calculated and depicted in Fig.~\ref{M87obs2} as function of $a$ and $l_0$ for a given $\alpha$ (=0.5, 1.0) taking inclination angle $\theta_0=90\si{\degree},17\si{\degree}$. The black curves denote the \textit{EHT} bound of $39 \mu$as, the $1\sigma$ bound of M87* $\theta_d$, and the region enclosed by it serves as the parameter space ($l_0-a$) consistent with the M87* observations.

\begin{figure*}
		\begin{tabular}{c c}
			\includegraphics[scale=0.6]{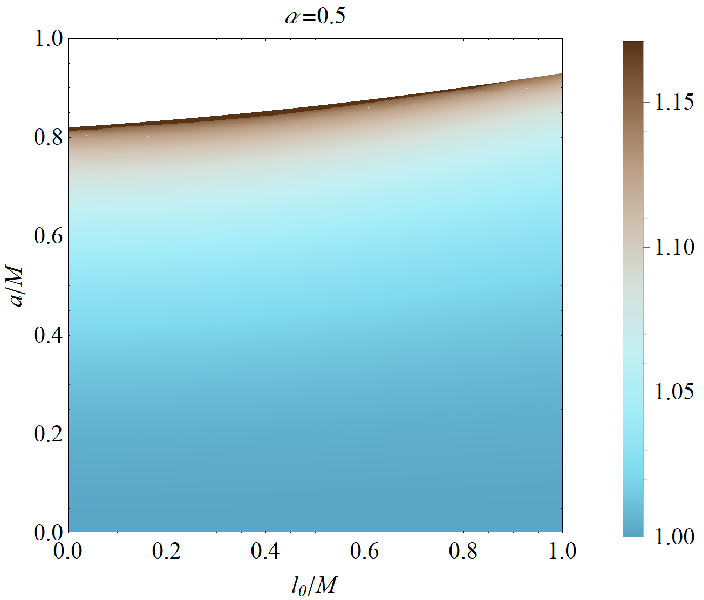}&
			\includegraphics[scale=0.6]{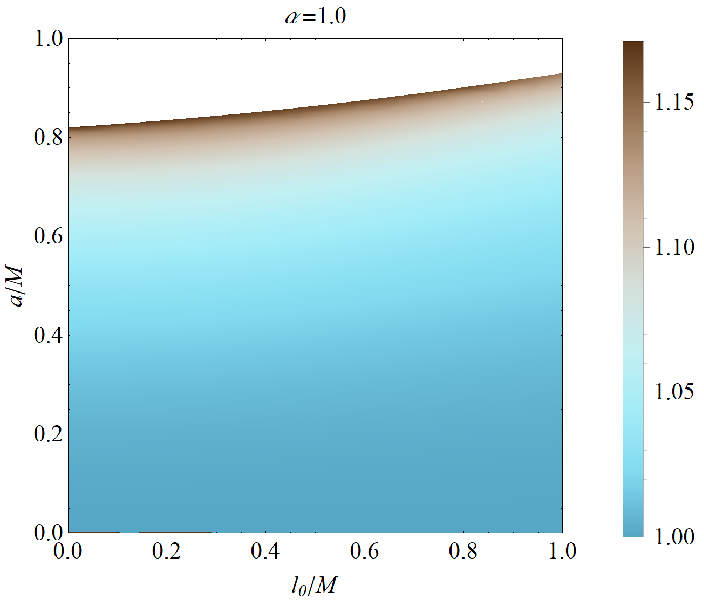}\\
				\includegraphics[scale=0.6]{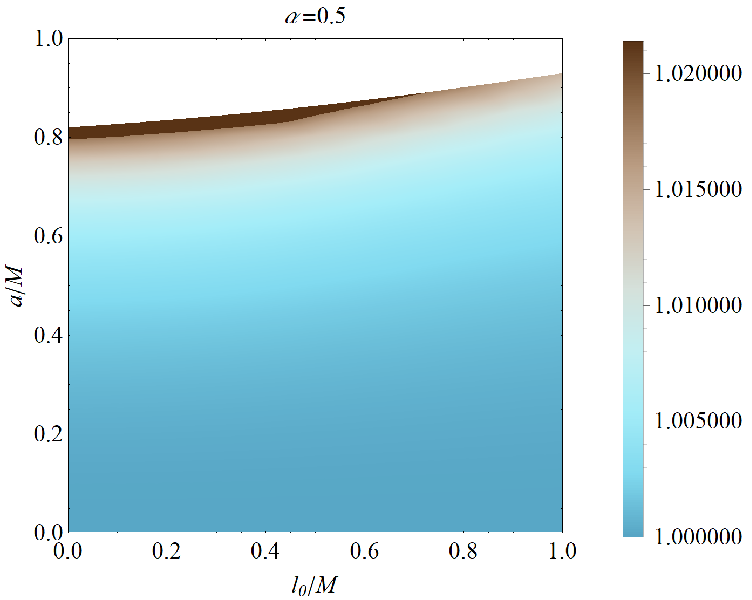}&
			\includegraphics[scale=0.6]{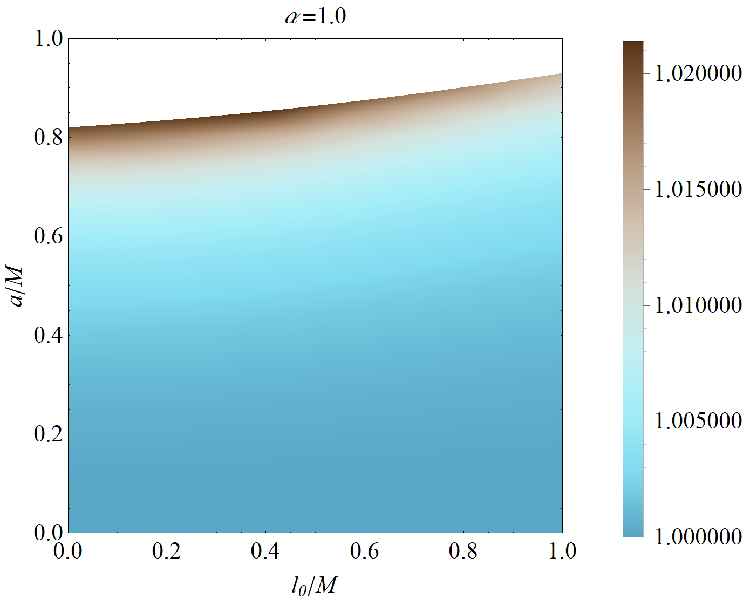}
		\end{tabular}	
	\caption{Axis ratio observable $D_x$ for hairy Kerr black hole shadows as a function of parameters ($a/M$ and $l_0/M$) at an inclination angle $\theta_0=90\si{\degree}$ (upper panel), $17\si{\degree}$ (lower panel) in agreement with \textit{EHT} results for M87*, $1 < D_{x} \lesssim 4/3$, is satisfied for entire parameter space ($a/M$ and $l_0/M$). The white region is forbidden for ($a/M$ and $l_0/M$).}
	\label{M87obs3}
\end{figure*}
The circular asymmetry in the M87* shadow can be redefined in terms of the axial ratio $D_x$ which is the ratio of major to minor diameter of the shadow \citep{Akiyama:2019cqa}. The axis ratio is given by \citep{Banerjee:2019nnj}
\begin{equation}
    D_x=\frac{\Delta Y}{\Delta X},
\end{equation}
should be in the range $1<D_x\lesssim 4/3$ in accordance with the \textit{EHT} observations of M87* \citep{Akiyama:2019cqa}. Indeed, $D_x$ is another way of defining $\Delta C$. It turns out that \textit{EHT} observed emission ring reconstructed in their images is close to circular with an axial ratio of 4:3, which indeed also corresponds to a $\Delta C \leq 0.1$ \citep{Akiyama:2019cqa} . The axis ratio is depicted as density plot in Fig.~\ref{M87obs3} and clearly, $1<D_x\lesssim 4/3$ is satisfied for the entire parameter space ($l_0$-$a$) of the hairy Kerr black holes and is thus remarkably consistent with \textit{EHT} images of M87* \citep{Akiyama:2019cqa}.  Thus, one can not rule out hairy Kerr black holes from the observational data of M87* black hole shadow. 
\section{Conclusions}\label{Sec7}
The no-hair theorem establishes the remarkable property of general-relativistic black holes that precisely three parameters, viz. mass, spin, and charge, uniquely determine their spacetimes and, hence, all of their properties \citep{Israel:1967za,Carter:1971zc,Hawking:1971vc}. The Kerr--Newman metric describe spacetimes of such black holes \citep{Newman:1965my}, which reduces to the Kerr metric \citep{Kerr:1963ud} for electrically neutral black holes or astrophysical black holes. Consequently, GR predicts that the Kerr metric describes all astrophysical black holes. Nonetheless, such black holes may not exist in a perfect vacuum due the presence of surrounding matter, like dark matter. The accretion disks could also alter the Kerr nature of the black hole. Non-Kerr spacetimes incorporate potential deviations from the Kerr metric by introducing additional deviation parameters and encompasses the Kerr metric if the deviation vanishes. Also, there are theoretical motivations (e.g. non-renormalizability and curvature singularities) to search for MoG. Further, the \textit{EHT} collaboration captured the image of supermassive black hole M87* at 1.3 $mm$ wavelength with an angular resolution of 20 $\mu\text{as}$, thereby opening a new window to test gravity in the strong-field regime.
The angular diameter of the shadow from \textit{EHT} observation of M87* is $\theta_d = 42 \pm 3 \mu as$  exhibiting a deviation from circularity $\Delta C \leq 10 \% $ and the axis ratio  $\lesssim 4/3$ consistent with the Kerr black hole's image as predicted by the GR \citep{Akiyama:2019cqa,Akiyama:2019bqs,Akiyama:2019eap}, but the observation did not say anything about most MoG or alternatives to the Kerr black hole. 
Motivated by this, we considered hairy Kerr black holes, which have additional parameters $\alpha$ and $l_0$  than the Kerr black hole, and the parameter $\alpha$ produces deviation from Kerr geometry but with a richer configuration. It is found that the hair parameter $l_0$, quantitatively influences the structure of the event horizon by reducing its radius significantly than that of the Kerr black hole, for a given $\alpha$, and the resulting increase in ergosphere area is thereby likely to have an impact on energy extraction, which has been investigated separately.  We have discussed the photons geodesics equations of motion, which are analytically solved in the first-order differential form. The possibility of using the available information of M87* shadow  to constraint hairy Kerr black holes, taking into account the surrounding dark matter, persuaded us to reconsider the shadow cast by the black holes. Observables, namely shadow radius $R_s$ and distortion $\delta_s$, characterize the shadow size and shape and estimate the black hole parameters ($a, l_0$). Besides, shadow observables, namely, area $A$ and oblateness $D$, are also used to explicitly determine the black hole parameters ($a, l_0$). Interestingly, the rotating hairy Kerr black holes, with the parameter $l_0$, cast smaller and more distorted shadows than the Kerr black holes, besides, the shadow size decreases, whereas the distortion increases with the decreasing value of parameter $l_0$. 

We highlight several other results obtained by considering M87* as a hairy Kerr black hole
\begin{enumerate}
\item The circularity deviation $\Delta C \leq 0.1$ is satisfied exhaustively for the entire ($l_0-a$) parameter space at both $\theta_0= 90\si{\degree}$ and $\theta_0=17\si{\degree}$ inclination angles.
\item The angular diameter satisfies $\theta_d= 42 \pm 3\mu$as within the $1 \sigma$ region over a finite ($l_0-a$) space. However, the accordant parameter space for $\theta_0=17\si{\degree}$ is smaller and more constricted than that at $\theta_0=90\si{\degree}$.
\item The axis ratio $D_x$ satisfies $1 < D_x \lesssim 4/3$ exhaustively for the entire ($l_0-a$) parameter space at both $\theta_0= 90\si{\degree}$ and $\theta_0=17\si{\degree}$ inclination angles.
\end{enumerate}
 Thus, our results constrain the hairy Kerr black holes parameters to ensure that the M87* shadow observations do not entirely rule out hairy Kerr black holes. The departure produced by the hairy Kerr black holes parameters $l_0$  and $\alpha$ for the M87* black hole shadow angular diameter is $\mathcal{O}(\mu$as). Hence,  it is difficult to distinguish the hairy black holes from the Kerr black hole, at least from the present resolution of astronomical observations like that of the \textit{EHT}, and we may have to wait for the next-generation EHT (ngEHT). 

The results presented here are the generalization of previous discussions, on the Kerr black hole, to a more general setting, following which the hairy Kerr black holes may allow studying the effect of surrounding dark matter on the Kerr black holes. Expectantly, the ngEHT, with highly improved techniques, would achieve the resolution. Moreover, the black hole shadow of  M87* due to EHT can also serve as a tool for distinguishing various models of theories of gravity thereby eventually shedding light on the nature of supermassive M87* and dark matter.  

In conclusion, 
 The \textit{EHT} observation has provided the first direct image of a supermassive black hole M87*. We have shown that the information gained from this observation place constraints on the hairy Kerr black holes.  It is difficult to distinguish the shadows of a Kerr and a hairy Kerr BH from the current \textit{EHT} observations; we find that the departure is $\mathcal{O}(\mu$as).  
Future observations of M87* by ngEHT can pin down the exact constraint and, in principle, may provide a signature of dark matter. 
Finally, our results, in the limit $\alpha \to 0$, reduced exactly  to  \emph{vis-$\grave{a}$-vis}  the Kerr black hole results.

\section*{Acknowledgements}
S.G.G. would like to thank SERB-DST for the ASEAN project IMRC/AISTDF/CRD/2018/000042. M.A. is supported by DST-INSPIRE Fellowship, DST, Govt. of India

\section*{Data Availability}

We have not generated any original data in the due course of this study, nor has any third-party data been analysed in this article.


\bibliographystyle{mnras}
\bibliography{example}
\bsp	
\label{lastpage}
\end{document}